\documentclass[
reprint,
bibnotes,
amsmath,amssymb,
aps,
prd
]{revtex4-2}

\usepackage{graphicx}
\usepackage{dcolumn}
\usepackage{enumitem}
\usepackage{bm}
\usepackage{float}
\usepackage[many]{tcolorbox}
\usepackage{shuffle}
\usepackage{environ}
\usepackage{physics}
\usepackage{tabularx}
\usepackage{diagbox}
\usepackage{subcaption}
\usepackage{multirow}

\usepackage{tikz}
\usetikzlibrary{calc}
\usetikzlibrary{shapes.geometric}
\usetikzlibrary{decorations.markings}
\usetikzlibrary{shapes.misc}
\tikzset{cross/.style={cross out, draw=black, minimum size=2*(#1-\pgflinewidth), inner sep=0pt, outer sep=0pt},
	cross/.default={2pt}}
\usepackage{CJK}

\begin{document}

\begin{CJK*}{UTF8}{}
\CJKfamily{gbsn}

\title{NLSM $\subset$ Tr$(\Phi^3)$}

\author{Nima Arkani-Hamed$^{1}$}
\email{arkani@ias.edu}
\author{Qu Cao(曹 趣)$^{2,3}$}
\email{qucao@zju.edu.cn}
\author{Jin Dong(董 晋)$^{2,4}$}
\email{dongjin@itp.ac.cn}
\author{Carolina Figueiredo$^{5}$}
\email{cfigueiredo@princeton.edu}
\author{Song He(何 颂)$^{2,6,7}$}
\email{songhe@itp.ac.cn}

\affiliation{$^{1}$School of Natural Sciences, Institute for Advanced Study, Princeton, NJ, 08540, USA \\
$^{2}$CAS Key Laboratory of Theoretical Physics, Institute of Theoretical Physics, Chinese Academy of Sciences, Beijing 100190, China \\
$^{3}$Zhejiang Institute of Modern Physics, Department of Physics, Zhejiang University, Hangzhou, 310027, China \\
$^{4}$School of Physical Sciences, University of Chinese Academy of Sciences, No.19A Yuquan Road, Beijing 100049, China \\
$^{5}$Jadwin Hall, Princeton University, Princeton, NJ, 08540, USA \\
$^{6}$School of Fundamental Physics and Mathematical Sciences, Hangzhou Institute for Advanced Study and ICTP-AP, UCAS, Hangzhou 310024, China\\
$^{7}$Peng Huanwu Center for Fundamental Theory, Hefei, Anhui 230026, P. R. China}

\begin{abstract}
Scattering amplitudes for the simplest theory of colored scalar particles -- the Tr($\Phi^3$) theory -- have recently been the subject of active investigations. In this letter we describe an unanticipated wider implication of this work: the Tr($\Phi^3$) theory secretly {\bf contains} Non-linear Sigma Model (NLSM) amplitudes to all loop orders. The NLSM amplitudes are obtained from Tr$(\Phi^3)$ amplitudes by a unique shift of kinematic variables. We show that this shifted kinematics produces amplitudes for a cubic theory with a linear term in potential, with extrema spontaneously breaking $U(N) \to U(N-k) \times U(k)$. The Goldstone amplitudes for this theory coincide with those of pions in the  $U(N) \times U(N) \to U(N)$ chiral Lagrangian to all orders in the planar limit. We also give a purely on-shell understanding of this correspondence, showing integrands defined by the kinematic shifts have the correct residues on poles and appropriately produce the Adler zero. Finally, we discuss how similar kinematic shifts produce certain infinite classes of mixed amplitudes of pions and Tr($\Phi^3$) scalars, most of which are not interpretable from the Lagrangian description. 
\end{abstract}
\maketitle
\end{CJK*}

\noindent {\bf Introduction.}---
The past few years have seen the discovery of rich combinatorial and geometric structures underlying scattering amplitudes of the simplest model of colored scalar particles, the Tr($\Phi^3$) theory. These include the kinematic associahedron for tree amplitudes~\cite{Arkani-Hamed:2017mur,Arkani-Hamed:2019vag}, and the ``curve-integral" formalism for amplitudes to all orders in the topological 't Hooft expansion~\cite{Arkani-Hamed:2023lbd, Arkani-Hamed:2023mvg}; the curve integrals are defined by a simple counting problem on any surface, which gives rise to striking ``binary geometries"~\cite{Arkani-Hamed:2019plo} both for ``stringy" amplitudes~\cite{Arkani-Hamed:2019mrd} as well as Tr($\Phi^3$) in the field-theory limit via ``tropicalization". 

The surprising connection between combinatorial geometry and scattering amplitudes began with the discoveries of the positive Grassmannian~\cite{Arkani-Hamed:2012zlh} and the amplituhedron in planar ${\cal N}=4$ super-Yang-Mills~\cite{Arkani-Hamed:2013jha}. Ever since this program has strived to more closely describe scattering amplitudes in the real world. The discovery of new combinatorial geometries without any supersymmetry in Tr($\Phi^3$) theory was an important step in this direction, but appeared to still miss the crucial richness and complexity associated with ``numerators'' of scattering amplitudes in realistic theories. 

But in~\cite{Zeros} we reported a major surprise: at tree-level, the Tr($\Phi^3$) amplitudes are in fact the $same$ as amplitudes for pions and gluons. Both the Non-linear Sigma Model (NLSM) and Yang-Mills (YM) amplitudes are contained in ``stringy" Tr$(\Phi^3)$ amplitudes by simple linear $\delta$eformation of the kinematic variables! This explains the surprising hidden zeros and new factorizations near zeros for tree amplitudes in all these theories ~\cite{Zeros}. 

This unity of colored scalars, pions, and gluons appears to extend to all loop orders. Gluons are obtained from ``scaffolded scalars'' which in turn come from stringy Tr$(\Phi^3)$ as explored at all loops in~\cite{Gluons}. As for pions, the connection can be described directly from field-theory Tr$(\Phi^3)$ amplitudes without referring to their ``stringy" definition.  The kinematic variables are the planar propagators $X_{i,j} = (p_i + p_{i+1} + \cdots p_{j-1})^2$. The $\delta$eformation of \cite{Zeros} are  $X_{o,o} \to X_{o,o}- \delta$ and $X_{e,e} \to X_{e,e}+\delta$; here ``$e$" and ``$o"$ refer to even and odd indices respectively. This kinematic shift preserves all the hidden zeros of the Tr$(\Phi^3)$ amplitudes. The leading order in the large $\delta\to \infty$ limit leads to NLSM tree amplitudes. 
The $\delta$ shift removes all the $X_{e,e},X_{o,o}$ poles, which are absent in the NLSM, while preserving the zeros, giving an object with correct poles/factorizations and the Adler zero~\cite{Adler:1964um}. 

In this letter, we show how this proposal extends to all loop orders. We begin by reviewing the story at tree-level in section I., and discuss the extension to all loops in sections II., seeing how in the $\delta\to \infty$ limit we obtain NLSM integrands with correct poles, factorizations, and forward-limits for single cuts. We will also see that they have the analog of the Adler zeros at loop level, with the integrand reducing to scaleless integrals in the soft limit. 

Then in section III. we give a Lagrangian derivation of the connection. We will show that the shifted Tr$(\Phi^3)$ theory corresponds precisely to a simple cubic Lagrangian implementing the spontaneous symmetry breaking pattern $U(N) \to U(N-k) \times U(k)$.
The color-ordered Goldstone amplitudes for this theory are computed by the low-energy/large $\delta$ expansion of the corresponding shifted Tr$(\Phi^3)$ amplitudes, to all orders in the topological expansion. An unusual feature of this simple cubic linear sigma model is that we expand around extrema of the cubic potential which includes local maxima and minima, so that there are heavy particles with negative mass$^2$ in the spectrum. Nonetheless, the low-energy integrands correctly describe the dynamics of the Goldstone particles enforced by the non-linear realization of the $U(N)/[U(N-k) \times U(k)]$ coset. Furthermore, the amplitudes for this non-linear sigma model coincide with the classic $U(N) \times U(N) \to U(N)$ symmetry breaking pattern for ``the'' non-linear sigma model for pions at leading order in the planar limit. 

Finally, a specific family of odd-point mixed amplitudes for pions and colored scalars naturally appeared as shifted Tr$(\Phi^3)$ in~\cite{Zeros}, associated with factorizations near zeros of NLSM amplitudes. In section IV. we extend this observation and initiate a systematic study of the mixed amplitudes that can and cannot be obtained from kinematic shifts of Tr$(\Phi^3)$. Only a small fraction of these amplitudes have a direct interpretation from our Lagrangian picture. 
\\ \\
\noindent {\bf I. NLSM from shifted Tr$(\Phi^3)$, tree-level.---}
\label{sec:nlsmfromshifted}
Let us first review the claim for the kinematic shifts giving the NLSM made in~\cite{Zeros}
\begin{equation*}
\tilde{X}_{i,j} = X_{i,j} + \delta_{i,j}\,, \quad 
 \delta_{e,e}=-\delta_{o,o}=\delta, \quad \delta_{o,e}=0.
\end{equation*}

Since the NLSM has only even-point interactions, the only propagators that occur in its amplitudes are the $X_{o,e}$, and so these are untouched. In addition, $\delta_{e,e}=-\delta_{o,o}$ ensures that all the non-planar variables remain unchanged
\begin{equation} \label{eq:ABHY}
c_{i,j}:=- 2 p_i \cdot p_j=\tilde{X}_{i,j}+\tilde{X}_{i{+}1, j{+}1}-\tilde{X}_{i,j{+}1}-\tilde{X}_{i{+}1, j}\,,  
\end{equation}
for $i<j{-}1$. As explained in~\cite{Zeros}, this means that the shifted amplitudes have all the hidden zeros found for Tr$(\Phi^3)$ amplitudes, which in turn imply the Adler zero of the NLSM amplitude after the $\delta$ shift. Further extracting the leading behavior at large $\delta$ yields 
\begin{equation}\label{tree}
\lim_{\delta\to \infty} \delta^{2n-2} A^{\delta}_{2n}(\tilde{X}_{i,j}) =A^{\rm NLSM}_{2n}(X_{i,j})\,.
\end{equation}

It may be physically more transparent to say we are working with $ X \ll \delta$. The leading amplitude is that of the NLSM with pion decay constant $f_\pi^2 = \delta$, and a tower of higher-dimension corrections. In multiplying by $\delta^{2n-2}$ and taking the $\delta \to \infty$ limit we get only the NLSM amplitudes, working in units with $f_\pi = 1$. It is non-trivial that the first non-vanishing order in the limit $\delta\to \infty$ is ${\cal O}(\delta^0)$, {\it i.e.} the contributions of ${\cal O}(\delta^{-m})$ for $m=1{-}2n/2, \cdots, -1$ all cancel, leaving us exactly with the mass dimension of $A_{2n}^{\rm NLSM}$. For the simplest case of $n=4$ starting from  $A_4^{\text{Tr}(\Phi^3)}=1/X_{1,3}+1/X_{2,4}$, and extracting limit \eqref{tree} gives
\begin{equation*}\begin{aligned}  \delta^2 \frac{X_{1,3}+X_{2,4}}{(X_{1,3}-\delta)(X_{2,4}+\delta)} \to -(X_{1,3}+X_{2,4})=A^{\rm NLSM}_4.\end{aligned}
\end{equation*}  

It is straightforward to prove \eqref{tree} for any $2n$ inductively. First of all, before we take $\delta \to \infty$, on any pole $X_{o,e}=0$, the shifted Tr$(\Phi^3)$ amplitude already factorizes into the product of two even-point amplitudes. In the large-$\delta$ limit they become corresponding NLSM amplitudes, thus we have shown the correct factorization on all physical poles. 
What remains to be shown is that it also has correct behavior at infinity, which concerns contact terms. It suffices to show that the limit has the Adler zero~\cite{Cheung:2014dqa, Cheung:2015ota}, namely it vanishes as any momentum becomes soft, {\it e.g.} $p_1\to 0$. This was shown in~\cite{Zeros} to follow from the ``skinny rectangle" zero: the shifted Tr$(\Phi^3)$ amplitude has a zero as $p_1\cdot p_i\to 0$ for $i=3,4,\cdots, n{-}1$, and it has no pole at $X_{1,3}=2 p_1\cdot p_2$ or $X_{2,n}=2 p_1\cdot p_n$. Thus the shifted amplitude and, in the $\delta\to \infty$ limit, the NLSM amplitude vanishes as $p_1\to 0$. 

We remark that tree amplitudes in the NLSM have been actively studied for over fifty years, with a resurgence of interest from the amplitudes program a decade ago ~\cite{Kampf:2013vha, Cachazo:2014xea}. The star of these studies has always been the Adler zero. By contrast, we have found more primitive zeros from which the Adler zero follows, directly connected to the shift of kinematic invariants that unifies NLSM and Tr$(\Phi^3)$ amplitudes.  
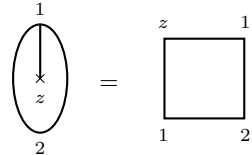
\begin{figure}[t]
\begin{equation*}
\begin{aligned}
\begin{tikzpicture}[scale=0.9]
    \draw[thick] (0,0) ellipse (0.4 and 0.8 );
    \draw (0,0) node[cross] {};
    \node at (0,-0.3) {\scriptsize $z$};
    \draw[thick] (0,0.8) node[above=0pt]{\scriptsize $1$};
    \draw[thick] (0,-0.8) node[below=0pt]{\scriptsize $2$};
    \draw[thick] (0,0)--(0,0.8);
\end{tikzpicture}
\end{aligned}\quad = \quad
\begin{aligned}
    \begin{tikzpicture}
    \node[regular polygon,minimum size = 1.5cm,regular polygon sides=4,draw,thick] (p) at (0,0) {};
	\draw[thick] (p.corner 1) node[above=0pt]{\scriptsize $1$};
	\draw[thick] (p.corner 2) node[above=0pt]{\scriptsize $z$};
	\draw[thick] (p.corner 3) node[below=0pt]{\scriptsize $1$};
	\draw[thick] (p.corner 4) node[below=0pt]{\scriptsize $2$};
    \end{tikzpicture}
\end{aligned}
\end{equation*}
    \caption{Single cut of one-loop 2-point NLSM with $X_{1,z_{1}}=0$ obtained from forward limit of 4-point tree.}
    \label{fig:1loop2ptsinglecut}
\end{figure}
\\ \\
\noindent {\bf II. NLSM from shifted Tr$(\Phi^3)$, loop level.---}
\label{sec:nlsmfromshifed} 
Let us first consider the one-loop extension of \eqref{tree}. In addition to the $X_{i,j}$'s, we now have loop propagators $X_{i,z}$ where $z$ is labelling the loop puncture. Denoting $X_{1,z}=\ell^2$ for some loop momentum $\ell$, we have $X_{i,z}=(\ell+p_1+ \cdots+ p_{i{-}1})^2$. Remarkably, we can apply \eqref{tree} with $z$ treated as another index, as long as we sum over the two possibilities of $z$ being even or odd, {\it i.e.}
\begin{equation}\label{1loop}
\lim_{\delta\to \infty} \delta^{2n} \left(A_{2n,1}^\delta (z~{\rm odd})+ A_{2n,1}^\delta(z~{\rm even})\right)=A_{2n,1}^{\rm NLSM}\,,
\end{equation}
where in each term we have the shifted kinematics also for the loop variables: $\tilde{X}_{i, z} = X_{i, z} + \delta_{i,z}$. As the simplest example, we consider the one-loop $2$-point Tr$(\Phi^3)$ integrand, which consists of a bubble and two tadpole diagrams, $A^{\text{Tr}(\Phi^3)}_{2,1}=1/(X_{1,z} X_{2,z})+1/(X_{1,z} X_{1,1})+1/(X_{2, z} X_{2,2})$, with $X_{i,i}$ denoting the tadpole variables. The $\delta\to \infty$ limit in \eqref{1loop}, yields
\begin{equation*}
\begin{aligned}
    A_{2,1}^{\delta}&=\frac 1 {(X_{1,z}-\delta)(X_{2,z})} +\frac 1 {(X_{1,z}-\delta)(X_{1,1}-\delta)}\\
    &+\frac 1 {(X_{2,z})(X_{2,2}+\delta)}+(1\leftrightarrow2, \delta \to -\delta)\\
    & \to \frac{1}{\delta^2} \left(2-\frac{X_{1,1}+X_{2,z}}{X_{1,z}}-\frac{X_{2,2}+X_{1,z}}{X_{2,z}}\right),
\end{aligned}
\end{equation*}
which is the one-loop $2$-point NLSM integrand $A^{\rm NLSM}_{2,1}$. This integrand is purely scaleless, but we can still check it has the correct single cut, giving us the forward limit of $4$-point NLSM tree (see Fig.~\ref{fig:1loop2ptsinglecut}):
\begin{equation*}
\underset{X_{1,z}=0}{\rm Res} A_{2,1}^{\rm NLSM}=-(X_{1,1}+X_{2,z})= A_{4}^{\rm NLSM}(1,z,1,2).
\end{equation*}

We obtain the one-loop $4$-point NLSM integrand in exactly the same way, and the result reads
\begin{equation} \label{eq:1loop4pt}
\begin{aligned}
    A^{\rm NLSM}_{4,1}&=\frac{\left(X_{2,z}+X_{1,3}\right) \left(X_{4,z}+X_{1,3}\right)}{X_{1,z} X_{3,z}}\\
    &+\frac{\left(X_{1,z}+X_{2,4}\right) \left(X_{3,z}+X_{2,4}\right)}{X_{2,z} X_{4,z}}\\
    &+(\text{scaleless}),
\end{aligned}
\end{equation}
where the first two terms are massive bubble integrals, and we omit the remaining terms since they are {\it scaleless} and therefore integrate to zero. (Here we omit terms with no $z$ dependence, or with only one $z$ in the denominator, {\it e.g.} $\frac{X_{2,z}}{X_{1,z}}$ which is scaleless by a change of variable).

The natural generalization of \eqref{1loop} to $L$ loops is to shift all the kinematic variables, including $X_{i,j}$, $X_{i,z_a}$ and $X_{z_a, z_b}$ for loop punctures $z_{\{a=1,\cdots, L\}}$. Crucially one must sum over all $2^L$ assignments of even/odd ``parity" for all $z_a$'s. The general claim is
\begin{equation}\label{loops}\lim_{\delta \to \infty} \sum^{2^L}_{{\rm even/odd}} \delta^{n{+}2L{-}2}A_{n, L}^{\delta}=A_{n,L}^{\rm NLSM}\,,
\end{equation}
{\it i.e.} the leading term in the expansion $\delta\to \infty$ of the shifted Tr($\Phi^3$) integrand, $A_{n,L}^\delta$, gives the planar $n$-point $L$-loop NLSM integrand. It is well known that there are ambiguities in the definition of loop integrands since one can always add terms that integrate to zero, but once more, one can check that \eqref{loops} gives an integrand with the correct factorizations and single cuts. 

All-loop planar Tr$(\Phi^3)$ integrands can be efficiently computed recursively ~\cite{recurtion} (see also~\cite{He:2018svj,Arkani-Hamed:2019vag}). Combined with our kinematic shifts, this allows us to efficiently compute NLSM tree amplitudes and loop integrands, which we have done at large multiplicity and through to three loops. 
For example, the two-loop $2$-point integrand reads~\footnote{Of course the entire $2$-point integrand is scaleless and by ``scaleless'' in the equation we suppress terms with $z_1$ or $z_2$ appearing in the denominator less than twice, {\it e.g.} $\frac{X_{1,z_2} X_{2,z_1}}{X_{1,z_1}^2 X_{z_1,z_2}}$}
\begin{equation*}
\begin{aligned}
    A^{\rm NLSM}_{2,2}&=\frac{\left(X_{1,z_{2}}+X_{2,z_{1}}\right)^{2} }{X_{1,z_{1}} X_{2,z_{2}}X_{z_{1},z_{2}}}+(z_{1}\leftrightarrow z_{2})\\
    &+(\text{scaleless}).
\end{aligned}
\end{equation*}
Let us briefly comment on the ``Adler zero" of loop amplitudes, and it is easy to see that only scaleless integrals survive in the soft limit (see~\cite{Bartsch:2022pyi} for recent studies). For example, since any unitarity (double) cut of a one-loop integrand equals the gluing of two tree amplitudes which, in turn, vanish in the soft limit, we can have only terms with vanishing unitarity cuts in the limit -- these are necessarily scaleless at one-loop. Explicitly, we find that under the soft limit {\it e.g.} $p_n\to 0$, the $n$-point integrand becomes a sum of (scaleless) $2$-point bubble integrals, which integrate to zero, times $(n{-}1)$-point tree-level ``mixed amplitude" with three $\Phi$'s; these mixed amplitudes appear as coefficients of Adler zero~\cite{Cachazo:2016njl} and in factorization near zeros~\cite{Zeros}, of $n$-point NLSM tree amplitude. It is also now understood that with an appropriate definition of ``surface soft'' limit, the Adler zero is an exact property of the full integrand \cite{circles}. 
\\ \\
\noindent {\bf III. The Lagrangian.}---
We now give a Lagrangian explanation for why this kinematic shift produces NLSM amplitudes. Let's consider a theory for an $N \times N$ scalar $\Psi$ with a general $U(N)$ invariant cubic potential
\begin{equation}
V(\Psi) = \frac{1}{3} g {\rm Tr} (\Psi^3) + \frac{1}{2} m^2 {\rm Tr}(\Psi^2) - c  {\rm Tr} (\Psi).
\end{equation}

The linear term means that the origin cannot be an extremum, which must satisfy $g \Psi^2 + m^2 \Psi - c = 0$. We can diagonalize $\Psi$, and the eigenvalues must satisfy this quadratic equation, which has two roots. Since the potential is cubic, one of the roots $r_-$ will correspond to a local minimum and the other $r_+$ to a local maximum of the potential. If $k$ of the eigenvalues are $r_+$ and $(N-k)$ are $r_-$, then we spontaneously break $U(N) \to U(N-k) \times U(k)$.
To identify the ``Goldstone'' and ``massive Higgs" modes, we expand around this extremum as $\Psi = {\rm diag}(r_+, \cdots, r_+;r_-,\cdots, r_-) + \Phi$, where we break up $\Phi$ to the obvious $k, (N-k)$ blocks as the fields $\phi_i^j, \tilde{\phi}_I^J$ for the $k \times k$, $(N-k) \times (N-k)$ diagonal blocks and $\chi_i^J,(\chi^\dagger)^i_J$ for the off-diagonal $k \times (N-k), (N-k) \times k$ blocks. Since we are expanding around the extremum, we will have only cubic and mass terms for these fields. The cubic term is just $g {\rm Tr} \Phi^3 = g [{\rm Tr} \phi^3 + 3 {\rm Tr} \phi \chi \chi^\dagger + 3 {\rm Tr} \tilde{\phi} \chi^\dagger \chi + {\rm Tr} \tilde{\phi}^3]$. The mass terms are interesting. Obviously $\chi, \tilde{\chi}$ are the massless Goldstone bosons. Since the double derivative of the potential is $2 g \Psi + m^2$, we have that $\phi$ has mass $m_\phi^2 = 2 g r_+ + m^2$ while $\tilde{\phi}$ has mass $m_{\tilde{\phi}}^2 = 2 g r_- + m^2$. But note that $m_{\phi}^2 + m_{\tilde \phi}^2 = 2 g (r_+ + r_-) + 2 m^2 = 0$ since the sum of the roots of the quadratic equation is $-m^2/g$. So we conclude that we have two massive ``Higgs" modes, with exactly equal and opposite mass$^2$. The necessity for this can of course also be seen directly from the $\phi, \tilde{\phi}, \chi, \chi^\dagger$ Lagrangian. Of course this Lagrangian linearly realizes the $U(N-k) \times U(k)$ symmetry, but  only with $m_\phi^2 = -m^2_{\tilde{\phi}}$ it is possible to add a constant shift term to the transformations of $\chi, \chi^\dagger$ to non-linearly realize the full $U(N)$ symmetry. 

This Lagrangian was described in the first version of our letter, (putting $N \to 2N$ and $k \to N$) as realizing the kinematic shift to the NLSM, as we will discuss just below. Here we add the observation that with $m^2_\phi = -m^2_{\tilde \phi}$ it non-linearly realizes the full $U(N)$ symmetry, and highlight its simple connection to standard spontaneous symmetry breaking starting from the $\Psi$ model. 

The fact that we have both positive and negative mass$^2$ ``massive Higgs modes'' is physically problematic but does not alter the fact that the amplitudes obtained from this theory at low momenta are the ones associated with the universal dynamics of the $U(N) \to U(N-k) \times U(k)$ coset. This is manifest at the level of loop integrands, but indeed for the purposes of loop integration, we can make $m_{\phi}^2 = - m_{\tilde{\phi}}^2$ purely imaginary so that there are no divergences associated with hitting the tachyonic poles. 

\begin{figure}
    \centering
\includegraphics[width=0.95\linewidth]{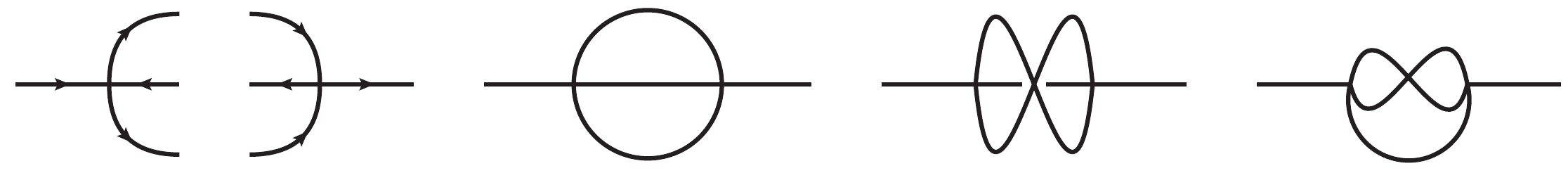}
    \caption{The 2-point 2-loop decorated graph of $\chi,\chi^\dagger$}(left), we can contract the internal lines to give the planar graph and non-planar graph(middle) but there are some non-planar graphs which could contribute to the NLSM are missing({\it e.g.} see right).
    \label{fig:nonplanar}
\end{figure}

Note that in looking at any color-trace factor for the Goldstone amplitudes, the $U(N-k) \times U(k)$ symmetry tells us they have to occur as Tr $(\chi_1 \chi_2^\dagger \chi_3 \chi_4^\dagger \cdots \chi_{2n-1} \chi_{2n}^\dagger)$. This contrasts with the pions of the NLSM, for which we have Tr$(\pi_1 \pi_2 \cdots \pi_{2n})$. Nonetheless, it is obvious that at tree-level the amplitudes are identical. The $\chi, \chi^\dagger$ amplitudes are only decorating those of the pions with additional ``in/out'' arrows, indicating whether the particle is a $\chi$ or $\chi^\dagger$. But the amplitudes factorize on poles in exactly the same way as pions, and have the same Adler zero; these properties are well-known to completely determine the trees and hence these amplitudes are the same. But this also tells us that the amplitudes for the two theories will be identical in the planar limit. We have just seen that the tree amplitudes and hence their respective contact terms, from which we are free to define the Lagrangian, are the same between the two theories. Now consider any planar diagram for the usual pion NLSM. Because it is planar, we can decorate all the (even-point) vertices with alternating in-out arrows, and this gives a corresponding diagram in the $\chi,\chi^\dagger$ theory. This holds for the leading $N$ contribution to amplitudes with any number of color trace factors. But beyond the planar limit, there is a difference. For instance consider a $2$-point function at 2-loops, with two quartic vertices with three internal lines each. For usual pions we can contract any of the three internal lines of one vertex with any of the other. But with in/out arrows, we can only contract an in arrow on one side with an out on the other, and so we are missing some of the contributions present for pions (see Fig.~\ref{fig:nonplanar}).

\begin{figure}
    \centering
    \includegraphics[width=0.55\linewidth]{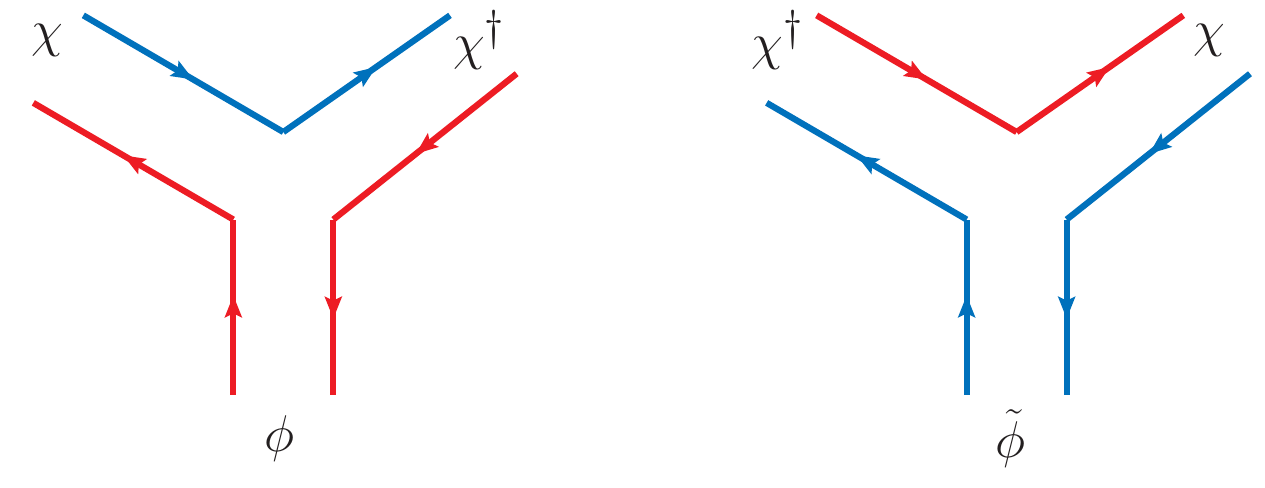}
    \caption{Vertices of ${\rm Tr} \phi \chi \chi^\dagger$ and ${\rm Tr} \tilde \phi \chi^\dagger \chi$}.
    \label{fig:vertex}
\end{figure}

Finally, it is easy to see that the kinematic shift of Tr$(\Phi^3)$ we have defined is doing nothing but computing the $(\chi \chi^\dagger \cdots)$ amplitudes from our Lagrangian. Consider any fatgraph/double-line notation cubic diagram involving $\phi, \tilde{\phi}, \chi, \tilde{\chi}$. We can color the $U(k)$ indices red and the $U(N-k)$ indices blue. External $\chi$ lines have an outgoing red arrow and an incoming blue one, oppositely for external $\chi^\dagger$ lines, and $\phi, \tilde{\phi}$ have in/outgoing red and blue lines respectively. Any propagator will be shifted by $-\delta$ if both its color lines are red, $+\delta$ if they are blue,  unshifted otherwise. A standard point that will be important in a moment is that our cubic interactions have a natural orientation: we have ${\rm Tr} \phi \chi \chi^\dagger$ and ${\rm Tr} \tilde \phi \chi^\dagger \chi$. Drawing the red/blue color lines for our interaction, there is a handedness, we can take the color lines to all turn left at the cubic vertex, but the same vertex where they are all turning right does not exist (see Fig.~\ref{fig:vertex}). This is the same as the basic fact that the surfaces associated with our cubic graphs are orientable, with this handedness of the vertices providing the orientation. 

Now for the kinematic shift. An elementary but crucial feature of ``surfaceology'' is that we can associate all propagators in diagrams with open curves on the surface/paths on the fatgraph, that are labelled by the external boundaries they touch, and in the case where the surfaces have puncture/there are closed internal color loops, by the punctures they spiral around. We simply walk out of the edge of a given propagator and turn left forever, in both direction, to end either on an external boundary or spiral around a  puncture. 

\begin{figure}
    \centering
    \includegraphics[width=0.95\linewidth]{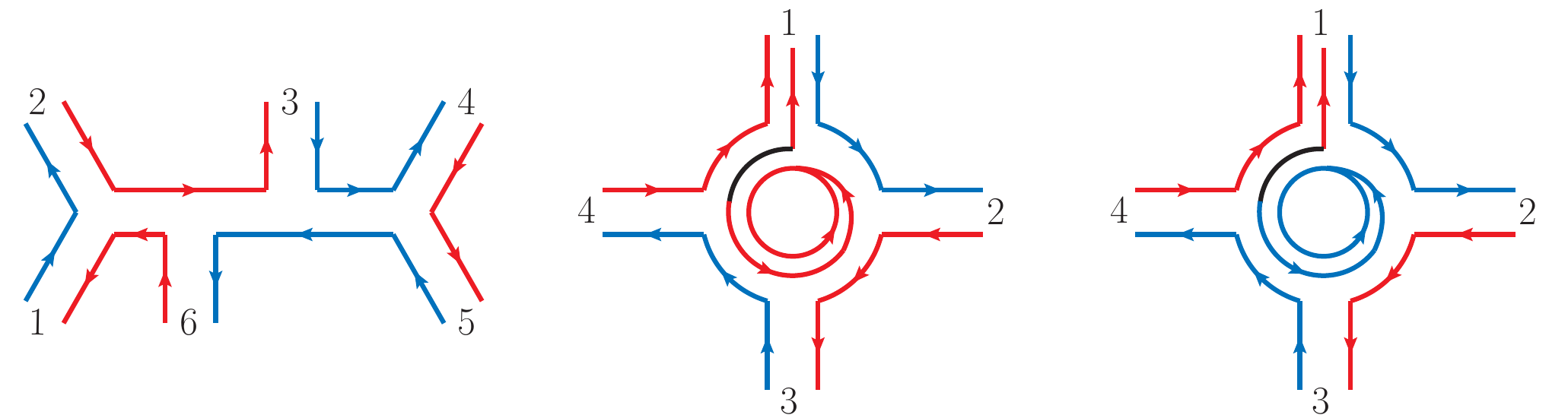}
    \caption{The 6-point tree diagram that contains negative shifted $X_{1,3}$, positive shifted $X_{4,6}$, unshifted $X_{3,6}$; and the 4-point 1-loop diagram with odd/even assignment to the loop puncture, where the path of the propagator $X_{1,z}$ goes out and spiral around the puncture with red, red/red, blue arrows, therefore should be shifted by $-\delta$/keep unshifted.}
    \label{fig:FeynDiagram}
\end{figure}

The point is that as we are walking out of the edge, we will be moving in the same direction as one of (red or blue) color arrows. The handedness of the color flow in our cubic interactions is then such that, if we always turn left, we will always continue to move in the same direction as an arrow of the same color. Let's first consider the case of surfaces with no punctures/closed internal color loops, so that every walk ends on an external boundary. If our initial edge has a red and a blue arrow, then when we walk out in one direction we will following a red arrow out, in the other a blue arrow out, and that tell us when we hit the boundary, we have to hit either a  $\chi$ or a $\phi$ on one end, and $\chi^\dagger$ or $\tilde{\phi}$ on the other. On the other hand, if our initial edge has both red arrows, we have to hit a $\chi, \phi$ on one end and an $\chi, \phi$ on the other, and similarly if our initial edge is all blue. Thus, if we label the external boundaries by the type of particle they are, we see that the kinematical variables are shifted as
\begin{equation}
    \begin{aligned}
        &X_{(\chi, \phi),(\chi, \phi)} \to X_{(\chi,\phi),(\chi,\phi)} - \delta, \\
        &X_{(\chi^\dagger, \tilde{\phi}), (\chi^\dagger, \tilde{\phi})} \to X_{(\chi^\dagger, \tilde{\phi}), (\chi^\dagger, \tilde{\phi})} + \delta,\\
        &X_{(\chi, \phi), (\chi^\dagger, \tilde{\phi})} \to X_{(\chi, \phi), (\chi^\dagger, \tilde{\phi})}.
    \end{aligned}
\end{equation}  

This kinematic shift of Tr$(\Phi^3)$ describes all the amplitudes in the theory. If we restrict to purely Goldstone scattering as Tr$(\chi_1 \chi^\dagger_2 \cdots)$, then since all the external $\chi$'s are odd and $\chi^\dagger$'s are even, this is the $X_{o,o} \to X_{o,o} - \delta, X_{e,e} \to X_{e,e} + \delta$ shift (see Fig.~\ref{fig:FeynDiagram} for examples). 

\begin{figure}
    \centering
    \includegraphics[width=0.45\linewidth]{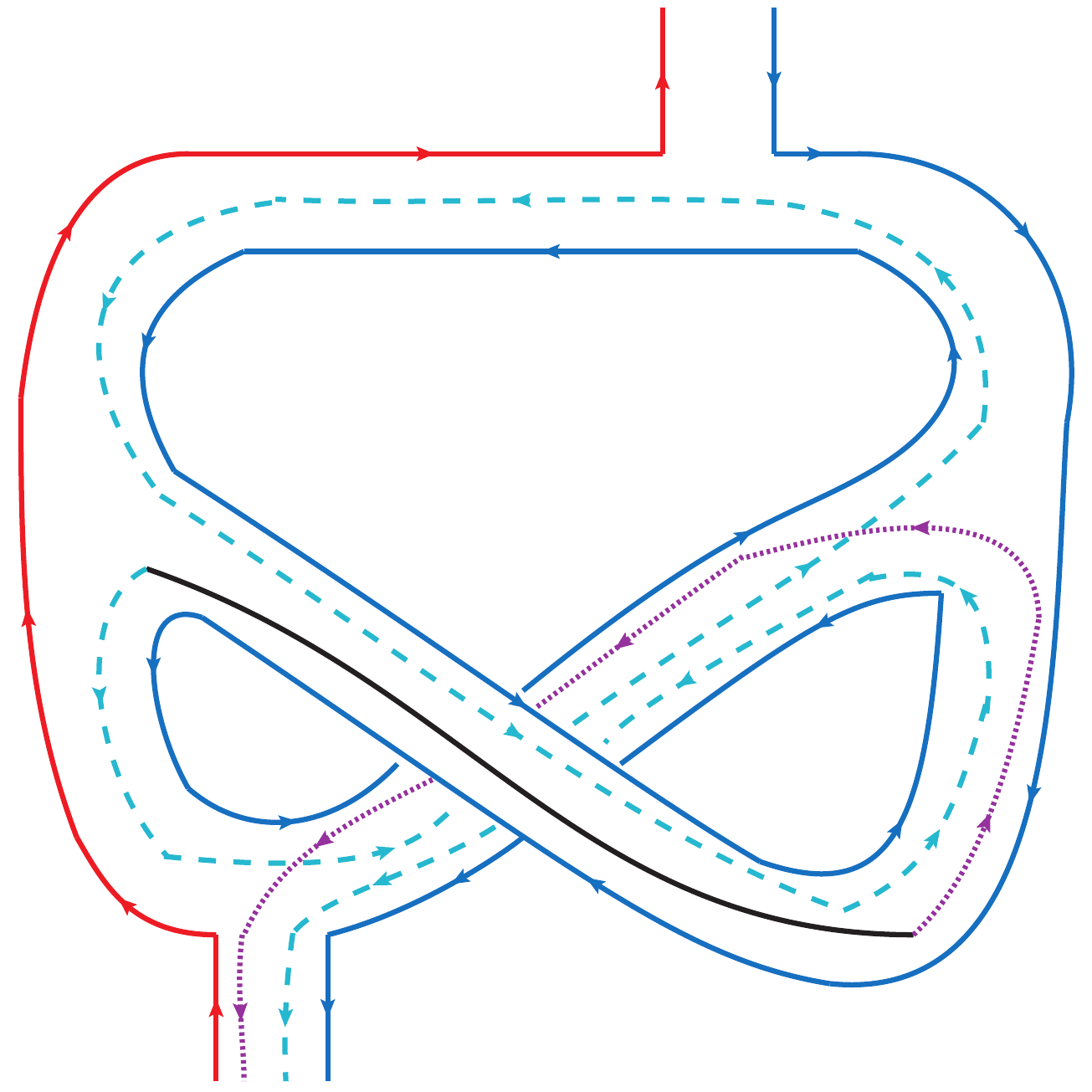}
    \caption{The shift of $\tilde \phi$ propagator in 2-point 2-loop non-planar example. The path of the $\tilde \phi$ propagator, denoted by black line with blue dashed or purple dotted arrows on each side inside the fat-graph. Each side extends outward with a corresponding external blue arrow, both of which terminate at the external $\chi^\dagger$. As a result, the corresponding propagator should be shifted by $+\delta$.}
    \label{fig:2loopFeyn}
\end{figure}

The case with punctures/closed internal color lines works in exactly the same way. Since we are summing over all colors, we must sum over the cases where the color lines are red and blue. The case where the puncture color is red is just as though we assign the puncture to be $(\chi, \phi)$-like, blue $(\tilde \chi, \tilde{\phi})$-like. This is simply the ``even/odd'' assignment above for punctures, and summing over red/blue internal colors simply sums over all such assignments.
Fig.~\ref{fig:FeynDiagram} and~\ref{fig:2loopFeyn} for $4$-point $1$-loop and $2$-point $2$-loop non-planar examples illustrate explicitly how this method allows us to determine the identity of internal propagator lines from the external particle labels. 

Obviously the general kinematic shift 
\begin{equation}
    \begin{aligned}
        &X_{(\chi, \phi),(\chi, \phi)} \to X_{(\chi,\phi),(\chi,\phi)} + m_\phi^2, \\
        &X_{(\chi^\dagger, \tilde{\phi}), (\chi^\dagger, \tilde{\phi})} \to X_{(\chi^\dagger, \tilde{\phi}), (\chi^\dagger, \tilde{\phi})} + m_{\tilde{\phi}}^2,\\
        &X_{(\chi, \phi), (\chi^\dagger, \tilde{\phi})} \to X_{(\chi, \phi), (\chi^\dagger, \tilde{\phi})} + m_\chi^2,
    \end{aligned}
\end{equation}   
produces amplitudes for our cubic Lagrangian with general masses for $\phi, \tilde \phi$ and $\chi$. As one application we can give both $\phi, \tilde \phi$ positive masses, integrating them out produces a Tr $(\chi^\dagger \chi \chi^\dagger \chi)$ theory with subleading corrections, and so this gives us an all-order description of this quartic theory. 

The fact that we can realize a non-linear sigma model via spontaneous symmetry breaking from a cubic Lagrangian is not in itself surprising. We can even realize the standard NLSM in this way, indeed with the same ingredients we have already seen above. Consider a cubic potential of the form $g$ Tr$(\phi \chi \chi^\dagger)$ + $\frac{1}{2} M^2$ Tr $(\phi^2)$ + $m^2$ Tr $(\chi^\dagger \chi)$, where these are all $N \times N$ matrices. This is just our cubic Lagrangian, with general masses for $\phi, \chi, \tilde{\phi}$ and for simplicity we are ignoring $\tilde{\phi}$ by sending its mass to infinity. This potential has a $U(N) \times U(N)$ symmetry. But if $M^2, m^2$ are either both positive or both negative, it has an extremum which spontaneously breaks the $U(N) \times U(N)$ to the diagonal $U(N)$. Indeed if $M^2$ is large, we can first integrate out $\phi$ and this generated a $-\frac{g^2}{M^2}$ Tr $(\chi^\dagger \chi)^2$ potential as just mentioned above.  If $M^2<0$ this is a positive quartic potential, and so if $m^2<0$ we have the classic Mexican hat potential, and the $\chi$ vev will spontaneously break $U(N) \times U(N)$ to the diagonal. The same happens if $M^2>0$, then we get a negative quartic coupling and so if $m^2>0$ we get an upside-down Mexican hat potential, and we can still expand around the extremum to get the same symmetry breaking pattern. It is amusing that our kinematic shift directly generates the amplitudes for this theory, expanded around the symmetric vacuum. But we can also expand around the symmetry breaking vacuum as $\phi = -\frac{m^2}{g} + \psi, \chi = \frac{m M}{g} + \zeta$ to obtain the cubic theory $g {\rm Tr}(\psi \zeta \zeta^\dagger) + \frac{1}{2} M^2 {\rm Tr} \psi^2 + (m M) {\rm Tr} \psi (\zeta + \zeta^\dagger)$. As usual the imaginary part of $\zeta$ are the massless Goldstones. If we further diagonalize the mass mixing between $\psi$ and the real part of $\zeta$, we get an $U(N)$ invariant cubic Lagrangian with a pattern of masses and interactions that non-linearly realizes the full $U(N) \times U(N)$ symmetry.
However, it is not trivial to obtain the amplitudes from this new cubic Lagrangian as a kinematic shift of Tr$(\Phi^3)$ theory. This is the extra feature of our simple cubic realization of $U(N) \to  U(N-k) \times U(k)$ that makes it remarkable --- the fact that we can so easily obtain the amplitudes for a seemingly more complicated theory from simple shifts of Tr$(\Phi^3)$, made possible by the special feature of being able to unambiguously determine the identity of inner propagator particles species from the external labels of the curves. 
\\ \\
\noindent {\bf IV. Mixed Amplitudes.---}
\label{sec:mixed amp}
We now discuss ``mixed amplitudes" of colored scalars $\Phi$ with a $\mu$Tr($\Phi^3$) coupling, interacting with the NLSM pions with a current-current interaction. At Lagrangian level, this gives derivative interactions between two $\Phi$'s and any even number of pions $\pi$. 
The color-ordered amplitudes are specified by a cyclically ordered pattern of $\Phi$'s and $\pi$'s, for instance, $(\Phi_1 \pi_2 \Phi_3 \pi_4 \Phi_5)$ at five points. 
Subsets of these mixed amplitudes are intimately connected to the pure NLSM, appearing as the coefficient multiplying the Adler zero of soft pions~\cite{Cachazo:2016njl}.

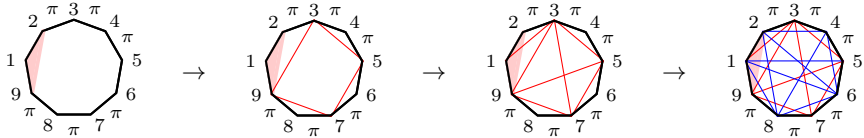
\begin{figure*}
	\centering
	\begin{equation*}
		\begin{aligned}
			&\begin{aligned}
				\begin{tikzpicture}[scale=1.1]
					\node[regular polygon,minimum size = 1.3 cm,regular polygon sides=9,draw,thick] (p) at (0,0) {};
					\draw (p.corner 1) node[above=-2pt]{\scriptsize $3$};
					\draw (p.corner 2) node[above left=-2pt]{\scriptsize $2$};
					\draw (p.corner 3) node[left=0pt]{\scriptsize $1$};
					\draw[thick] (p.corner 4) node[left=0pt]{\scriptsize $9$};
					\draw (p.corner 5) node[below left=-2pt]{\scriptsize $8$};
					\draw (p.corner 6) node[below right=-2pt]{\scriptsize $7$};
					\draw (p.corner 7) node[right=0pt]{\scriptsize $6$};
					\draw (p.corner 8) node[right=0pt]{\scriptsize $5$};
					\draw (p.corner 9) node[above right=-2pt]{\scriptsize $4$};
					\node at (110:0.75) {\scriptsize $\pi$};
					\node at (70:0.75) {\scriptsize $\pi$};
					\node at (30:0.75) {\scriptsize $\pi$};
					\node at (-90:0.75) {\scriptsize $\pi$};
					\node at (-47:0.75) {\scriptsize $\pi$};
					\node at (-135:0.75) {\scriptsize $\pi$};
					\fill[fill=red!20!white] (p.corner 2)--(p.corner 3)--(p.corner 4)--cycle ;
                    \draw[red!20!white,line width=0.5mm] (p.corner 2)+(0.025,0.01)--($(0.02,0.01)+(p.corner 4)$);
					
					\node[regular polygon,minimum size = 1.3 cm,regular polygon sides=9,draw,thick] (q) at (0,0) {};
				\end{tikzpicture}
			\end{aligned}
			\quad \to \quad 
			\begin{aligned}
				\begin{tikzpicture}[scale=1.1]
					\node[regular polygon,minimum size = 1.3 cm,regular polygon sides=9,draw,thick] (p) at (0,0) {};
					\draw (p.corner 1) node[above=-2pt]{\scriptsize $3$};
					\draw (p.corner 2) node[above left=-2pt]{\scriptsize $2$};
					\draw (p.corner 3) node[left=0pt]{\scriptsize $1$};
					\draw[thick] (p.corner 4) node[left=0pt]{\scriptsize $9$};
					\draw (p.corner 5) node[below left=-2pt]{\scriptsize $8$};
					\draw (p.corner 6) node[below right=-2pt]{\scriptsize $7$};
					\draw (p.corner 7) node[right=0pt]{\scriptsize $6$};
					\draw (p.corner 8) node[right=0pt]{\scriptsize $5$};
					\draw (p.corner 9) node[above right=-2pt]{\scriptsize $4$};
					\node at (110:0.75) {\scriptsize $\pi$};
					\node at (70:0.75) {\scriptsize $\pi$};
					\node at (30:0.75) {\scriptsize $\pi$};
					\node at (-90:0.75) {\scriptsize $\pi$};
					\node at (-47:0.75) {\scriptsize $\pi$};
					\node at (-135:0.75) {\scriptsize $\pi$};
					
					\fill[fill=red!20!white] (p.corner 2)--(p.corner 3)--(p.corner 4)--cycle ;
                        \draw[red!20!white,line width=0.5mm] (p.corner 2)+(0.025,0.01)--($(0.02,0.01)+(p.corner 4)$);
					
					\draw[red] (p.corner 1)--(p.corner 4); \draw[red] (p.corner 1)--(p.corner 8);
					\draw[red] (p.corner 6)--(p.corner 8); \draw[red] (p.corner 6)--(p.corner 4);

					\node[regular polygon,minimum size = 1.3 cm,regular polygon sides=9,draw,thick] (q) at (0,0) {};
				\end{tikzpicture}
			\end{aligned}
			\quad \to \quad 
			\begin{aligned}
				\begin{tikzpicture}[scale=1.1]
					\node[regular polygon,minimum size = 1.3 cm,regular polygon sides=9,draw,thick] (p) at (0,0) {};
					\draw (p.corner 1) node[above=-2pt]{\scriptsize $3$};
					\draw (p.corner 2) node[above left=-2pt]{\scriptsize $2$};
					\draw (p.corner 3) node[left=0pt]{\scriptsize $1$};
					\draw[thick] (p.corner 4) node[left=0pt]{\scriptsize $9$};
					\draw (p.corner 5) node[below left=-2pt]{\scriptsize $8$};
					\draw (p.corner 6) node[below right=-2pt]{\scriptsize $7$};
					\draw (p.corner 7) node[right=0pt]{\scriptsize $6$};
					\draw (p.corner 8) node[right=0pt]{\scriptsize $5$};
					\draw (p.corner 9) node[above right=-2pt]{\scriptsize $4$};
					\node at (110:0.75) {\scriptsize $\pi$};
					\node at (70:0.75) {\scriptsize $\pi$};
					\node at (30:0.75) {\scriptsize $\pi$};
					\node at (-90:0.75) {\scriptsize $\pi$};
					\node at (-47:0.75) {\scriptsize $\pi$};
					\node at (-135:0.75) {\scriptsize $\pi$};
					
					\fill[fill=red!20!white] (p.corner 2)--(p.corner 3)--(p.corner 4)--cycle ;
                        \draw[red!20!white,line width=0.5mm] (p.corner 2)+(0.025,0.01)--($(0.02,0.01)+(p.corner 4)$);
					
					\draw[red] (p.corner 1)--(p.corner 4); \draw[red] (p.corner 1)--(p.corner 8);
					\draw[red] (p.corner 6)--(p.corner 8); \draw[red] (p.corner 6)--(p.corner 4);
					
					\draw[red] (p.corner 1)--(p.corner 6); \draw[red] (p.corner 8)--(p.corner 4);
					\draw[red] (p.corner 1)--(p.corner 3);

					\node[regular polygon,minimum size = 1.3 cm,regular polygon sides=9,draw,thick] (q) at (0,0) {};
				\end{tikzpicture}
			\end{aligned}
			\quad \to \quad 
			\begin{aligned}
				\begin{tikzpicture}[scale=1.1]
					\node[regular polygon,minimum size = 1.3 cm,regular polygon sides=9,draw,thick] (p) at (0,0) {};
					\draw (p.corner 1) node[above=-2pt]{\scriptsize $3$};
					\draw (p.corner 2) node[above left=-2pt]{\scriptsize $2$};
					\draw (p.corner 3) node[left=0pt]{\scriptsize $1$};
					\draw[thick] (p.corner 4) node[left=0pt]{\scriptsize $9$};
					\draw (p.corner 5) node[below left=-2pt]{\scriptsize $8$};
					\draw (p.corner 6) node[below right=-2pt]{\scriptsize $7$};
					\draw (p.corner 7) node[right=0pt]{\scriptsize $6$};
					\draw (p.corner 8) node[right=0pt]{\scriptsize $5$};
					\draw (p.corner 9) node[above right=-2pt]{\scriptsize $4$};
					\node at (110:0.75) {\scriptsize $\pi$};
					\node at (70:0.75) {\scriptsize $\pi$};
					\node at (30:0.75) {\scriptsize $\pi$};
					\node at (-90:0.75) {\scriptsize $\pi$};
					\node at (-47:0.75) {\scriptsize $\pi$};
					\node at (-135:0.75) {\scriptsize $\pi$};
					
					\fill[fill=red!20!white] (p.corner 2)--(p.corner 3)--(p.corner 4)--cycle ;
				    \draw[red!20!white,line width=0.5mm] (p.corner 2)+(0.025,0.01)--($(0.02,0.01)+(p.corner 4)$);
        
					\draw[red] (p.corner 1)--(p.corner 4); \draw[red] (p.corner 1)--(p.corner 8);
					\draw[red] (p.corner 6)--(p.corner 8); \draw[red] (p.corner 6)--(p.corner 4);
					
					\draw[red] (p.corner 1)--(p.corner 6); \draw[red] (p.corner 8)--(p.corner 4);
					\draw[red] (p.corner 1)--(p.corner 3);
					
					\draw[blue] (p.corner 2)--(p.corner 9); \draw[blue] (p.corner 9)--(p.corner 7);
					\draw[blue] (p.corner 7)--(p.corner 5); \draw[blue] (p.corner 5)--(p.corner 2);
					\draw[blue] (p.corner 2)--(p.corner 7); \draw[blue] (p.corner 9)--(p.corner 5);
					
					\draw[blue] (p.corner 3)--(p.corner 7); \draw[blue] (p.corner 3)--(p.corner 8);
					\draw[blue] (p.corner 3)--(p.corner 5);
					
					\node[regular polygon,minimum size = 1.3 cm,regular polygon sides=9,draw,thick] (q) at (0,0) {};
				\end{tikzpicture}
			\end{aligned}
		\end{aligned}
	\end{equation*}
	\caption{An illustration of the four-step algorithm for a 9-point mixed amplitude. Step 1, identifying the shaded $\Phi$ region. Step 2, building the inner-gon. Step 3, shifted chords inside the inner-gon and its boundaries. Step 4, shifting the remaining forbidden chords in blue.}
\label{fig:allstep}
\end{figure*}
This model has two coupling constants $\mu$ and $\frac{1}{f_\pi}$. We will be interested in the leading low-energy amplitude, with the most powers of $\mu$. The leading Feynman diagrams for $n_{\Phi}$ $\Phi$'s and $2m$ $\pi$'s are easy to describe: they are $n_\Phi$-point Tr($\Phi^3$) diagram decorated with $\pi$'s coming off $\Phi$ lines. These amplitudes are characterized by the following simple properties: amplitudes with an odd number of $\pi$'s as well as those with a single $\Phi$  vanish; they have an Adler zero as any pion momentum is taken to zero; the four-point amplitude for two $\Phi$'s and two $\pi$'s is the same as for four $\pi$'s; finally for the leading low-energy units, we have factorization on $\Phi$ poles and $\pi$-poles only when one of the left or right factors is a pure pion amplitude. As a consequence, the power-counting for an amplitude with $n_\Phi$ $\Phi$'s and $2 m$ $\pi$'s is simply $X^{3 - n_\Phi}$. 

Let us now discuss how and to what extent these mixed amplitudes are related to shifted Tr($\Phi^3$) amplitudes. The reason to suspect such a connection was seen in \cite{Zeros}, where a certain class of mixed amplitudes appears in the near-zero factorizations of NLSM amplitudes. The pattern was one with three $\Phi$'s, with two of them adjacent, and separated from the third $\Phi$ with any number (possibly zero) of $\pi$'s. As obvious from their origin (and as we will understand more generally in a moment), these amplitudes can be obtained as certain shifts of Tr($\Phi^3$) amplitudes that do not preserve {\it all} the zeros, but only those associated with the ``skinny rectangle" zeros of the pion legs.  It is then natural to ask whether further patterns of mixed $\Phi/\pi$ amplitudes can be obtained by more general shifts of Tr($\Phi^3$) amplitudes. 

Indeed we have found rich connections between mixed amplitudes and shifts of Tr($\Phi^3$), which fall into three categories: ($i$) infinite classes of amplitudes generalizing those encountered in near-zero-factorization, for which the mixed amps are given by shifts of Tr($\Phi^3$) that preserve the skinny rectangle zeros for the pions; ($ii$) infinite classes obtainable by shifts of Tr($\Phi^3$) without any obvious connection to preserving skinny rectangle zeros, motivated by a general algorithm we explain below;  $(iii)$ mixed amplitudes that are {\bf impossible} to obtain from shifts of Tr$(\Phi^3)$ amplitudes, coinciding with cases where the general algorithm encounters an obstruction. 

We now describe these three classes of phenomena, leaving a full exploration of these questions to future work. We begin with $(i)$ and the simplest mixed amplitude, with two $\Phi$'s and any number $2m$ of $\pi$'s. It is clear from diagrams that these are identical to $(2 + 2m)$-pt NLSM amplitudes, but let us understand this in the language of shifts. To begin with we note that the poles that are forbidden for this amplitude are all the $X_{e,e},X_{o,o}$ just as in the NLSM (note that the importance of single $\Phi$ amplitudes vanishing for this to be true). We can then shift all the $X_{e,e},X_{o,o}$ of the $(2 + 2m)$ point Tr ($\Phi^3$) amplitude by arbitrary amounts to remove the poles while still preserving the skinny rectangle zeros. It is a pretty fact that imposing just $2m$ of these zeros (rather than all $(2 + 2m)$ of them) actually suffices to force all the shifts to be those of the full NLSM. 

But having done this, we can generalize to non-trivial examples where the two $\Phi$' are ``blown up" into blocks of adjacent $\Phi$'s. If the two $\Phi$'s are adjacent to begin with this gives us the pattern $[(\Phi \cdots \Phi)(\pi \cdots \pi)]$, while if they are non-adjacent we have the more general $[(\Phi \cdots \Phi)(\pi \cdots \pi)(\Phi \cdots \Phi)(\pi \cdots \pi)]$. All the $X_{i,j}$'s in the regions bounded by the $\Phi$ blocks are unshifted, while the shifts for all the chords touching the initial $\Phi$ before blow-up are inherited by all the rest of the $\Phi$'s in the block. This clearly preserves the skinny rectangle zeros for the $\pi$'s and only shifts away the unwanted poles of the blown-up $\Phi/\pi$ configuration. 
An illustration of this is given in appendix~\ref{app:exapmle}. 

We next present an algorithm to determine the shifts that produce the pattern of $\Phi/\pi$ for a given mixed amplitude from ($ii$). This algorithm has been checked for all the mixed amplitudes through to 8 points. The first obstruction to implementing the algorithm appears in some 9-point examples, and it is precisely for these cases that the mixed amplitude \textbf{cannot} be obtained via shifting. 

To obtain a given mixed amplitude from Tr$(\Phi^3)$ we want to shift \textbf{all} chords that cannot appear as factorization channels in the mixed case. This we can determine just by checking that the left and right factors of a given chord are allowed. The non-trivial part is determining the relative sign of the shifts. To do this, we consider the momentum $n$-gon of the corresponding mixed amplitude and we use the following four-step procedure (see Fig.~\ref{fig:allstep}):
\begin{enumerate}[topsep=3pt,itemsep=0pt,partopsep=1ex,parsep=1ex]
    \item Color all the subregions of the $n$-gon bounded by adjacent $\Phi$'s. Inside these regions, all chords are allowed since they come from Tr($\Phi^3$) interactions. As for the chords bounding these regions if lower amplitudes are not allowed, we should color them in red~\footnote{Note that this is only the case if all the $\Phi$'s are adjacent to each other.}. 
    \item Considering the effective problem where each $\Phi$-region is replaced by a single $\Phi$, color in red the chords, $X_{i,i+2}$, forming an inner-gon~\footnote{Note that if the effective problem is a $4$-point the inner-gon is simply a chord.}. If the effective problem is an odd-point, the inner-gon does not close and we require that the side left open corresponds to a $\pi$ surrounded by $\Phi$'s. In some cases this might not be possible, as we will show below, corresponding to a case where the amplitude cannot be obtained by shifts. If in 1. the chord bounding the $\Phi$ region is colored in red, then it should be part of the inner-gon.
    \item Consider all chords living inside the inner-gon and for those in which lower-point factors are not allowed color them in red. Similarly, consider all the boundaries of the inner-gon that contain a $\Phi$-region and a $\pi$. Color in red all chords that bound the $\pi$ and any number of $\Phi$'s, which are not allowed. 
    \item All the remaining chords for which one of the lower-point factors is not allowed, should be colored in blue. In most cases, this procedure automatically guarantees that for each point $i$ inside a $\pi$ region, we have that all the colored chords $X_{i,j}$ have a different color than that of $X_{i-1,i+1}$. However, there are some chords in this last step that need to be colored red for this to be the case. 
\end{enumerate}

Finally, by shifting all the red chords with the opposite sign of the blue chords, we get the desired mixed amplitude. One surprise we ran into while understanding the algorithm is the existence of \textbf{flat-directions}: in some cases, the amplitude is independent of some $\delta_{i,j}$'s, which must only be non-zero to shift away unwanted poles. 

We give a number of examples of this algorithm in action in appendix~\ref{app:exapmle}, including an infinite class of mixed amplitudes of the form $[(\Phi_1) \pi_1 (\Phi_2) \pi_2 \cdots (\Phi_{2m}) \pi_{2m}]$, where each $(\Phi_i)$ stands for a region of adjacent $\Phi$'s. 

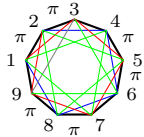
\begin{figure}
    \centering
\begin{tikzpicture}[scale=1.1]
\node[regular polygon,minimum size = 1.3 cm,regular polygon sides=9,draw,thick] (p) at (0,0) {};
	\draw (p.corner 1) node[above=-2pt]{\scriptsize $3$};
	\draw (p.corner 2) node[above left=-2pt]{\scriptsize $2$};
	\draw (p.corner 3) node[left=0pt]{\scriptsize $1$};
	\draw[thick] (p.corner 4) node[left=0pt]{\scriptsize $9$};
	\draw (p.corner 5) node[below left=-2pt]{\scriptsize $8$};
	\draw (p.corner 6) node[below right=-2pt]{\scriptsize $7$};
	\draw (p.corner 7) node[right=0pt]{\scriptsize $6$};
	\draw (p.corner 8) node[right=0pt]{\scriptsize $5$};
 	\draw (p.corner 9) node[above right=-2pt]{\scriptsize $4$};
	 \node at (110:0.75) {\scriptsize $\pi$};
	 \node at (150:0.75) {\scriptsize $\pi$};
      \node at (30:0.75) {\scriptsize $\pi$};
      \node at (-7:0.75) {\scriptsize $\pi$};
      \node at (-90:0.75) {\scriptsize $\pi$};
      \node at (-135:0.75) {\scriptsize $\pi$};
      
    \draw[blue] (p.corner 9)--(p.corner 2);
    \draw[blue] (p.corner 9)--(p.corner 7);	
    \draw[blue] (p.corner 5)--(p.corner 7);	
    \draw[blue] (p.corner 5)--(p.corner 3);	
    
    \draw[red] (p.corner 1)--(p.corner 3);
    \draw[red] (p.corner 1)--(p.corner 8);
    \draw[red] (p.corner 6)--(p.corner 8);
    \draw[red] (p.corner 6)--(p.corner 4);
    \draw[red] (p.corner 2)--(p.corner 4);

    \draw[green] (p.corner 2)--(p.corner 8);
    \draw[green] (p.corner 5)--(p.corner 8);
    \draw[green] (p.corner 2)--(p.corner 5);

    \draw[green] (p.corner 1)--(p.corner 7);
    \draw[gray] (p.corner 1)--(p.corner 4);
    \draw[green] (p.corner 7)--(p.corner 4);

    \draw[green] (p.corner 3)--(p.corner 9);
    \draw[green] (p.corner 6)--(p.corner 9);
    \draw[green] (p.corner 6)--(p.corner 3);
\end{tikzpicture}
    \caption{An obstruction for $9$-point mixed amplitude $[\pi_1 \pi_2 \Phi_3 \pi_4 \pi_5 \Phi_6 \pi_7 \pi_8 \Phi_9]$.}
    \label{exceptional case}
\end{figure}

Let us finally consider $(iii)$, mixed amplitudes that can not be generated by shifts of Tr($\Phi^3$). The simplest example is the nine-point $[\Phi_1 \pi_2 \pi_3 \Phi_4 \pi_5 \pi_6 \Phi_7 \pi_8 \pi_9]$. We suppose this amplitude can be obtained by shifting from Tr$(\Phi^3)$ with any shift parameters, and obtain a contradiction. This amplitude has a pole when $X_{1,4} \to 0$, where it factorizes into a product of lower amplitudes one of which is the $4$-point associated with the quadrilateral $(1,2,3,4)$. We know that to correctly reproduce this amplitude, we must have the shifts $\delta_{1,3} = -\delta_{2,4}$. Similarly, we have a factorization when $X_{2,5} \to 0$, where one factor is $(2,3,4,5)$, which tells us we must have $\delta_{2,4} = -\delta_{3,5}$. In this way we can go around all the way to the factorization where $X_{2,8} \to 0$, and this determines $\delta_{1,3} = -\delta_{2,4} = \delta_{3,5} = -\delta_{4,6} = \delta_{5,7} = -\delta_{6,8} = \delta_{7,9}=-\delta_{1,8}=\delta_{2,9}$. We consider the final factorization with $X_{3,9} \to 0$ which demands $\delta_{1,3} = -\delta_{2,9}$, but we have forced instead that $\delta_{1,3} = + \delta_{2,9}$, which leads to a contradiction (see Fig.~\ref{exceptional case}). 

Last but not least, we also find that similar shifts can produce some mixed amplitudes at loop level as well. For simplicity, we study one infinite class of mixed amplitudes, namely  $A_{n,L}(\pi_1\pi_2\Phi_3\cdots\Phi_n)$ in appendix~\ref{app:Loop}. Already here we can see how it is a simple generalization of what we observed at tree level. We leave the question about more general configurations for future work.

It is interesting to ask whether any of these mixed amplitudes can be related to amplitudes of the $\phi, \tilde{\phi}, \chi, \chi^\dagger$ Lagrangian. Clearly, some interpretation will be needed, since unlike the case for $\chi,\chi^\dagger$ scattering where the particles are massless, our $\phi, \tilde{\phi}$ are massive and their relation to the $\phi$'s of the mixed amplitudes is not immediate. There are some very simple examples where a connection can be made however. Suppose we consider the $\phi \phi \chi \chi^\dagger$ amplitude. Using our rules this is simply $\frac{1}{X_{1,3} - \delta} + \frac{1}{X_{2,4}}$. But since the particle $\phi$ has a big mass, if we want to imagine that the $\chi$'s have low momentum, then while $X_{1,3} = (p_{1,\phi} + p_{2,\phi})^2 = (p_{3, \chi} + p_{4, \chi^\dagger})^2$ is small, $X_{2,4} = (p_{2,\phi} + p_{3,\chi})^2 = p_2^2 + $ small = $\delta + $ small. So it makes sense to expand $X_{2,4} = \delta + x_{2,4}$, and now expand for $X_{1,3}$ and $x_{2,4}$ both small. This gives exactly the 4-point NLSM amplitude which is also the mixed amplitude for $(\phi \phi \pi \pi)$. The same thing happens for an amplitude for two adjacent $\phi$'s and any number of $\chi \chi^\dagger$, and also with three adjacent $\phi$'s and any number of $\chi, \chi^\dagger$. But already the mixed amplitude for $(\phi \pi \phi \pi)$, where the $\phi$'s are separated, can not be interpreted in any straightforward way. They could only be related to the $(\phi \chi \tilde \phi  \chi^\dagger)$ amplitude, that our rules give as $\frac{1}{X_{1,3}} + \frac{1}{X_{2,4}}$. But now because $\phi, \tilde \phi$ have opposite mass$^2$, it is simply impossible to take any soft limit for the $\chi, \tilde \chi$, and no obvious way to relate their kinematics to those of massless $(\phi \pi \phi \pi)$ in the mixed amplitudes. 
\\ \\
\noindent {\bf Outlook.---}
We have seen an all-loop connection between the scattering amplitudes of Tr$(\Phi^3)$ and the non-linear sigma model via a simple kinematical shift. This is both striking given that these theories are seemingly utterly different, and also useful, since the recent progress in all-loop formulations of Tr$(\Phi^3)$ theory immediately gives us the same all-loop information about NLSM. There are many obvious avenues for further investigation opened up by these results, and we end by discussing some of them. 

Loop integrands for the NLSM are not unique; depending (up to terms that integrate to zero) on the choice of Lagrangian representation. It would be interesting to look for a parametrization of the NLSM Lagrangian that produces our integrand. It is also natural to expect that our integrand has an especially nice soft behavior. Indeed, the soft limits of our all-loop integrands can always be written as simple scaleless integrals dressed with ``universal" amplitudes~\cite{jarainprogress}, while the most natural notion of ``surface soft'' limit gives an exact Adler zero even at the level of the integrand \cite{circles}. 

More generally, what is special about our unusual UV completion of the NLSM relative to e.g. ordinary linear sigma completions is that our amplitudes enjoy the hidden zero/factorization properties as exact statements even for finite $\delta$, not just at low-energies. We can see this already at four points, where our amplitude is $\frac{1}{X_{1,3} - \delta} + \frac{1}{X_{2,4} + \delta}$, which has the simplest ``hidden zero''  when $X_{1,3} + X_{2,4} = 0$, for any $\delta$. Of course because this amplitude isn't cyclically symmetric at finite $\delta$ we can't interpret it as a pion amplitude, but we can symmetrize over the sign of $\delta$ to get a symmetric amplitude, that obviously does have this zero. By contrast the usual linear sigma model completion with massive Higgses has amplitude $\frac{1}{X_{1,3} - \delta} + \frac{1}{X_{2,4} - \delta} + \frac{2}{\delta}$. This gives the same low-energy NLSM amplitude, but does not vanish when $(X_{1,3} + X_{2,4}) = 0$. As expected, due to the tachyons the four-derivative and higher Wilson coefficients of the low-energy effective theory for our amplitudes will violate positivity bounds. We can see this explicitly by expanding the (sign of $\delta$ symmetrized) four-point amplitude, the coefficients of $X_{1,3}^2 = s^2$ is zero, and this should be strictly positive. It would be interesting to explore whether there are physically healthy UV completions of the NLSM (unlike ours with tachyons) that also enjoy this zero. 
They can certainly pass the first checks of consistency with positivity bounds: if we take the 4-point amplitude to be an infinite expansion in powers of $u = -(X_{1,3} + X_{2,4})=-(s+t)$, we ensure the hidden zero, and it is well-known that polynomials in $u$ are trivially consistent with a positive partial wave expansion.


Beyond pure Goldstone amplitudes, we have seen that large families of ``mixed'' amplitudes can also be obtained from kinematic shifts of Tr$(\Phi^3)$. As we alluded to above our simple cubic Lagrangian is simply related to a very small number of these, but most still don't have a Lagrangian interpretation. It would be interesting to better understand Lagrangian interpretations for the full set of amplitudes, and the connections between them.

Moving from integrands to full amplitudes, given that loop-integrated amplitudes of Tr($\Phi^3$) can be computed via the curve-integral formalism of~\cite{Arkani-Hamed:2023lbd}, by the kinematic shift also gives new ``tropical'' representations for integrated NLSM amplitudes as well \cite{tropLag, circles}. 

The discovery of a connection between Tr$(\Phi^3)$, the NLSM and Yang-Mills theory was motivated by the experimental observation that the hidden zeroes and factorizations of Tr$(\Phi^3)$ theory are shared by pions and gluons. In this letter we have seen that the connections between the NLSM and shifted Tr$(\Phi^3)$ theory has a simple Lagrangian interpretation, but the zeroes and factorizations, that both motivated and made the connection obvious and natural to begin with, remains a magical feature of the Tr($\Phi^3$) theory with no known Lagrangian origin. It would be obviously be interesting to look for such an understanding. 

Compared to Yang-Mills amplitudes which can only be obtained from the fully ``stringy" Tr($\Phi^3$) amplitudes~\cite{Gluons}, NLSM amplitudes are much simpler as they directly follow from the field-theoretic Tr($\Phi^3$) amplitude.  But it is also interesting that the NLSM appears universally from the point of view of deformed stringy amplitudes: as discussed in \cite{Zeros} the NLSM always arises in the low-energy theory for any non-integer $\delta$. For large $\delta$ close to integers, the NLSM is intriguingly UV completed by a tower of particles including not only gluons but also higher spin states, going to infinite spin as $\delta \to \infty$. We expect further discoveries are awaiting us in this new world of connections between colored scalars, pions, gluons, and their particle/stringy amplitudes. 
\\ \\
\noindent {\bf Acknowledgements.---}
It is our pleasure to thank C. Bartsch, K. Kampf and J. Trnka for stimulating discussions. The work of N.A.H. is supported by the DOE (Grant No. DE-SC0009988), by the Simons Collaboration on Celestial Holography, and by the Carl B. Feinberg cross-disciplinary program in innovation at the IAS. The work of Q.C. is supported by the National Natural Science Foundation of China under Grant No. 123B2075. The work of C.F. is supported by FCT/Portugal
(Grant No. 2023.01221.BD). The work of S.H. has been supported by the National Natural Science Foundation of China under Grant No. 12225510, 11935013, 12047503, 12247103, and by the New Cornerstone Science Foundation through the XPLORER PRIZE.

\appendix
\section{Examples of tree-level mixed amplitudes from shifts}
\label{app:exapmle}
\begin{figure}[H]
    \centering
\begin{equation*}
\begin{aligned}
\begin{tikzpicture}[scale=1.1]
\node[regular polygon,minimum size = 1.3cm,regular polygon sides=6,draw,thick,rotate=-30] (p) at (0,0) {};

\draw[thick] (p.corner 1) node[right]{\scriptsize $3$};
\draw[thick] (p.corner 2) node[above]{\scriptsize $2$};
\draw[thick] (p.corner 3) node[left]{\scriptsize $1$};
\draw[thick] (p.corner 4) node[left]{\scriptsize $n$};
\draw[thick] (p.corner 5) node[below]{\scriptsize $\ldots$};
\draw[thick] (p.corner 6) node[right]{\scriptsize $4$};
\node at (120:0.7) {\scriptsize $\pi$};
\node at (60:0.7) {\scriptsize $\pi$};

\fill[fill=red!20!white] (p.corner 3)--(p.corner 1)--(p.corner 6)--(p.corner 5)--(p.corner 4)--cycle ;

\draw[thick] (p.corner 3)--(p.corner 4);
\draw[thick,dashed] (p.corner 4)--(p.corner 5);
\draw[thick,dashed] (p.corner 5)--(p.corner 6);
\draw[thick] (p.corner 1)--(p.corner 6);

\draw[red] (p.corner 1)--(p.corner 3);
\end{tikzpicture}
\end{aligned}
\; \to \;
\begin{aligned}
\begin{tikzpicture}[scale=1.1]
\node[regular polygon,minimum size = 1.3cm,regular polygon sides=6,draw,thick,rotate=-30] (p) at (0,0) {};

\draw[thick] (p.corner 1) node[right]{\scriptsize $3$};
\draw[thick] (p.corner 2) node[above]{\scriptsize $2$};
\draw[thick] (p.corner 3) node[left]{\scriptsize $1$};
\draw[thick] (p.corner 4) node[left]{\scriptsize $n$};
\draw[thick] (p.corner 5) node[below]{\scriptsize $\ldots$};
\draw[thick] (p.corner 6) node[right]{\scriptsize $4$};
\node at (120:0.7) {\scriptsize $\pi$};
\node at (60:0.7) {\scriptsize $\pi$};

\fill[fill=red!20!white] (p.corner 3)--(p.corner 1)--(p.corner 6)--(p.corner 5)--(p.corner 4)--cycle ;

\node at (0,0) {\scriptsize ${\color{blue} \cdots}$};

\draw[thick] (p.corner 3)--(p.corner 4);
\draw[thick,dashed] (p.corner 4)--(p.corner 5);
\draw[thick,dashed] (p.corner 5)--(p.corner 6);
\draw[thick] (p.corner 1)--(p.corner 6);

\draw[red] (p.corner 1)--(p.corner 3);
\draw[blue] (p.corner 2)--(p.corner 6);
\draw[blue] (p.corner 2)--(p.corner 4);
\end{tikzpicture}
\end{aligned}
\end{equation*}
    \caption{The algorithm for the shifts for  $[\pi_1\pi_2\Phi_3\Phi_4\ldots \Phi_n]$.}
    \label{fig:mixed2pion}
\end{figure}
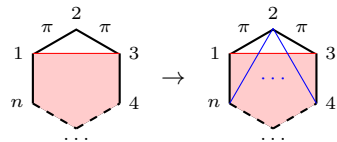

We begin with illustrating our algorithm for shifts with the simplest non-trivial case of two adjacent pions, $[\pi_1 \pi_2 \Phi_3  \cdots \Phi_n]$. The forbidden poles are $X_{1,3}$ and $X_{2,i}$ for $i=4,5, \cdots, n$, and for the first step the $\Phi$-region is found to be $(1,3, 4, \cdots, n)$ (bounded by the chord $(13)$); we color $X_{1,3}$ red and since the effective problem is simply $[\pi \pi \Phi]$ where no chord is needed inside, thus all we need is to color all the chords $X_{2,i}$ blue (see Fig.~\ref{fig:mixed2pion}). 

\begin{figure}
    \centering
\begin{equation*}
\begin{aligned}
\begin{tikzpicture}[scale=1.1]
\node[regular polygon,minimum size = 1.3 cm,regular polygon sides=8,draw,thick] (p) at (0,0) {};
	\draw[thick] (p.corner 1) node[above right=-2pt]{\scriptsize $3$};
	\draw[thick] (p.corner 2) node[above left=-2pt]{\scriptsize $2$};
	\draw[thick] (p.corner 3) node[left=0pt]{\scriptsize $1$};
	\draw[thick] (p.corner 4) node[left=0pt]{\scriptsize $8$};
	\draw[thick] (p.corner 5) node[below left=-2pt]{\scriptsize $7$};
	\draw[thick] (p.corner 6) node[below right=-2pt]{\scriptsize $6$};
	\draw[thick] (p.corner 7) node[right=0pt]{\scriptsize $5$};
	\draw[thick] (p.corner 8) node[right=0pt]{\scriptsize $4$};
    \node at (140:0.75) {\scriptsize $\pi$};
    \node at (-40:0.75) {\scriptsize $\pi$};
    \fill[fill=red!20!white] (p.corner 2)--(p.corner 1)--(p.corner 8)--(p.corner 7)--cycle ;
    \fill[fill=red!20!white] (p.corner 3)--(p.corner 4)--(p.corner 5)--(p.corner 6)--cycle ;
\node[regular polygon,minimum size = 1.3 cm,regular polygon sides=8,draw,thick] (q) at (0,0) {};
\end{tikzpicture}
\end{aligned}
\; \to \;
\begin{aligned}
\begin{tikzpicture}[scale=1.1]
\node[regular polygon,minimum size = 1.3 cm,regular polygon sides=8,draw,thick] (p) at (0,0) {};
	\draw[thick] (p.corner 1) node[above right=-2pt]{\scriptsize $3$};
	\draw[thick] (p.corner 2) node[above left=-2pt]{\scriptsize $2$};
	\draw[thick] (p.corner 3) node[left=0pt]{\scriptsize $1$};
	\draw[thick] (p.corner 4) node[left=0pt]{\scriptsize $8$};
	\draw[thick] (p.corner 5) node[below left=-2pt]{\scriptsize $7$};
	\draw[thick] (p.corner 6) node[below right=-2pt]{\scriptsize $6$};
	\draw[thick] (p.corner 7) node[right=0pt]{\scriptsize $5$};
	\draw[thick] (p.corner 8) node[right=0pt]{\scriptsize $4$};
	\node at (140:0.75) {\scriptsize $\pi$};
	\node at (-40:0.75) {\scriptsize $\pi$};
    \fill[fill=red!20!white] (p.corner 2)--(p.corner 1)--(p.corner 8)--(p.corner 7)--cycle ;
    \fill[fill=red!20!white] (p.corner 3)--(p.corner 4)--(p.corner 5)--(p.corner 6)--cycle ;
    \draw[red] (p.corner 2)--(p.corner 6);
    \draw[red] (p.corner 2)--(p.corner 5);
    \draw[red] (p.corner 2)--(p.corner 4);

    \draw[red] (p.corner 6)--(p.corner 1);
    \draw[red] (p.corner 6)--(p.corner 8);
\node[regular polygon,minimum size = 1.3 cm,regular polygon sides=8,draw,thick] (q) at (0,0) {};
\end{tikzpicture}
\end{aligned}
\; \to \;
\begin{aligned}
\begin{tikzpicture}[scale=1.1]
\node[regular polygon,minimum size = 1.3 cm,regular polygon sides=8,draw,thick] (p) at (0,0) {};
	\draw[thick] (p.corner 1) node[above right=-2pt]{\scriptsize $3$};
	\draw[thick] (p.corner 2) node[above left=-2pt]{\scriptsize $2$};
	\draw[thick] (p.corner 3) node[left=0pt]{\scriptsize $1$};
	\draw[thick] (p.corner 4) node[left=0pt]{\scriptsize $8$};
	\draw[thick] (p.corner 5) node[below left=-2pt]{\scriptsize $7$};
	\draw[thick] (p.corner 6) node[below right=-2pt]{\scriptsize $6$};
	\draw[thick] (p.corner 7) node[right=0pt]{\scriptsize $5$};
	\draw[thick] (p.corner 8) node[right=0pt]{\scriptsize $4$};
	\node at (140:0.75) {\scriptsize $\pi$};
	\node at (-40:0.75) {\scriptsize $\pi$};
    \fill[fill=red!20!white] (p.corner 2)--(p.corner 1)--(p.corner 8)--(p.corner 7)--cycle ;
    \fill[fill=red!20!white] (p.corner 3)--(p.corner 4)--(p.corner 5)--(p.corner 6)--cycle ;
    \draw[blue] (p.corner 3)--(p.corner 7);
    \draw[blue] (p.corner 3)--(p.corner 8);
    \draw[blue] (p.corner 3)--(p.corner 1);

    \draw[blue] (p.corner 7)--(p.corner 4);
    \draw[blue] (p.corner 7)--(p.corner 5);

    \draw[red] (p.corner 2)--(p.corner 6);
    \draw[red] (p.corner 2)--(p.corner 5);
    \draw[red] (p.corner 2)--(p.corner 4);

    \draw[red] (p.corner 6)--(p.corner 1);
    \draw[red] (p.corner 6)--(p.corner 8);

    \draw[blue] (p.corner 4)--(p.corner 1);
    \draw[blue] (p.corner 5)--(p.corner 1);
    \draw[blue] (p.corner 4)--(p.corner 8);
    \draw[blue] (p.corner 5)--(p.corner 8);
    
\node[regular polygon,minimum size = 1.3 cm,regular polygon sides=8,draw,thick] (q) at (0,0) {};
\end{tikzpicture}
\end{aligned}
\end{equation*}
\caption{The algorithm for the shifts for $[\pi_1\Phi_2\Phi_3\Phi_4\pi_5\Phi_6\Phi_7\Phi_8]$.}
    \label{fig:mixed8pt}
\end{figure}
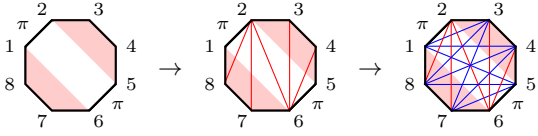

We next consider two non-adjacent pions, for example, $[\pi_1\Phi_2\Phi_3\Phi_4\pi_5\Phi_6\Phi_7\Phi_8]$. In step 1, we have two $\Phi$ regions which are shaded red in Fig.~\ref{fig:mixed8pt}. Then in step $2$, the effective problem is $[\pi \Phi \pi \Phi]$, so we color the chord $(2,6)$ in red. In step $3$, the chords $(2,7),(2,8),(3,6),(4,6)$ are not allowed for they bound the $\pi$ and some $\Phi$'s, which are in red. For the final step, the remaining forbidden factorization channels corresponding to chords $(i,j)$ with $i=3,4,5$ and $j=1,7,8$, are all colored in blue.

Similarly for the generic case with two non-adjacent $\pi$'s and $(n{-2})$ $\Phi$'s, $\pi_1 (\Phi) \pi_i (\Phi)$, we obtain the shift rule:
\begin{equation*}
\begin{aligned}
&{-}\delta_{1,i}={-}\delta_{1,j}={-}\delta_{i,k}=-\delta_{j,k}=\delta_{2,i{+}1}=\delta_{j,i+1}=\delta_{2,k}=\delta,
\end{aligned}
\end{equation*}
for $j=3,\cdots,i{-}1,$ and $ k=i{+2},\cdots,n$. 

\begin{figure}
    \centering
\begin{equation*}
\begin{aligned}
\begin{tikzpicture}[scale=1.1]
\node[regular polygon,minimum size = 1.3cm,regular polygon sides=8,draw,thick] (p) at (0,0) {};

\draw[thick] (p.corner 1) node[above right=-2pt]{\scriptsize $3$};
\draw[thick] (p.corner 2) node[above left=-2pt]{\scriptsize $2$};
\draw[thick] (p.corner 3) node[left=0pt]{\scriptsize $1$};
\draw[thick] (p.corner 4) node[left=0pt]{\scriptsize $8$};
\draw[thick] (p.corner 5) node[below left=-2pt]{\scriptsize $7$};
\draw[thick] (p.corner 6) node[below right=-2pt]{\scriptsize $6$};
\draw[thick] (p.corner 7) node[right=0pt]{\scriptsize $5$};
\draw[thick] (p.corner 8) node[right=0pt]{\scriptsize $4$};
\node at (140:0.75) {\scriptsize $\pi$};
\node at (90:0.75) {\scriptsize $\pi$};
\node at (0:0.75) {\scriptsize $\pi$};
\node at (40:0.75) {\scriptsize $\pi$};
\fill[fill=red!20!white] (p.corner 3)--(p.corner 7)--(p.corner 6)--(p.corner 5)--(p.corner 4)--cycle ;
\draw[red] (p.corner 1)--(p.corner 3);
\draw[red] (p.corner 1)--(p.corner 7);
\draw[red] (p.corner 3)--(p.corner 7);

\draw[blue] (p.corner 2)--(p.corner 4);
\draw[blue] (p.corner 2)--(p.corner 5);
\draw[blue] (p.corner 2)--(p.corner 6);
\draw[blue] (p.corner 2)--(p.corner 8);

\draw[blue] (p.corner 4)--(p.corner 8);
\draw[blue] (p.corner 5)--(p.corner 8);
\draw[blue] (p.corner 6)--(p.corner 8);
    
\node[regular polygon,minimum size = 1.3 cm,regular polygon sides=8,draw,thick] (q) at (0,0) {};
\end{tikzpicture}
\end{aligned}
\quad
\begin{aligned}
\begin{tikzpicture}[scale=1.1]
\node[regular polygon,minimum size = 1.3 cm,regular polygon sides=8,draw,thick] (p) at (0,0) {};
	\draw[thick] (p.corner 1) node[above right=-2pt]{\scriptsize $3$};
	\draw[thick] (p.corner 2) node[above left=-2pt]{\scriptsize $2$};
	\draw[thick] (p.corner 3) node[left=0pt]{\scriptsize $1$};
	\draw[thick] (p.corner 4) node[left=0pt]{\scriptsize $8$};
	\draw[thick] (p.corner 5) node[below left=-2pt]{\scriptsize $7$};
	\draw[thick] (p.corner 6) node[below right=-2pt]{\scriptsize $6$};
	\draw[thick] (p.corner 7) node[right=0pt]{\scriptsize $5$};
	\draw[thick] (p.corner 8) node[right=0pt]{\scriptsize $4$};
	\node at (140:0.75) {\scriptsize $\pi$};
	\node at (-40:0.75) {\scriptsize $\pi$};
	\node at (-140:0.75) {\scriptsize $\pi$};
	\node at (40:0.75) {\scriptsize $\pi$};
	\draw[blue] (p.corner 1)--(p.corner 3);
	\draw[blue] (p.corner 1)--(p.corner 7);
	\draw[blue] (p.corner 7)--(p.corner 5);
	\draw[blue] (p.corner 3)--(p.corner 5);

	\draw[red] (p.corner 2)--(p.corner 4);
	\draw[red] (p.corner 2)--(p.corner 8);
	\draw[red] (p.corner 6)--(p.corner 8);
	\draw[red] (p.corner 4)--(p.corner 6);
	
	\draw[blue] (p.corner 2)--(p.corner 7);
	\draw[blue] (p.corner 3)--(p.corner 6);
	\draw[blue] (p.corner 1)--(p.corner 4);
	\draw[blue] (p.corner 8)--(p.corner 5);
\end{tikzpicture}
\end{aligned}
\quad
\begin{aligned}
\begin{tikzpicture}[scale=1.1]
\node[regular polygon,minimum size = 1.3 cm,regular polygon sides=9,draw,thick] (p) at (0,0) {};
	\draw (p.corner 1) node[above=-2pt]{\scriptsize $3$};
	\draw (p.corner 2) node[above left=-2pt]{\scriptsize $2$};
	\draw (p.corner 3) node[left=0pt]{\scriptsize $1$};
	\draw[thick] (p.corner 4) node[left=0pt]{\scriptsize $9$};
	\draw (p.corner 5) node[below left=-2pt]{\scriptsize $8$};
	\draw (p.corner 6) node[below right=-2pt]{\scriptsize $7$};
	\draw (p.corner 7) node[right=0pt]{\scriptsize $6$};
	\draw (p.corner 8) node[right=0pt]{\scriptsize $5$};
    \draw (p.corner 9) node[above right=-2pt]{\scriptsize $4$};
	 \node at (110:0.75) {\scriptsize $\pi$};
	 \node at (150:0.75) {\scriptsize $\pi$};
      \node at (70:0.75) {\scriptsize $\pi$};
      \node at (-7:0.75) {\scriptsize $\pi$};
      \node at (-47:0.75) {\scriptsize $\pi$};
      \node at (-135:0.75) {\scriptsize $\pi$};
      
    \draw[red] (p.corner 4)--(p.corner 2); \draw[red] (p.corner 2)--(p.corner 9);
    \draw[red] (p.corner 9)--(p.corner 7);\draw[red] (p.corner 7)--(p.corner 5);
    \draw[red] (p.corner 7)--(p.corner 2);\draw[red] (p.corner 9)--(p.corner 4);
    
    \draw[blue] (p.corner 3)--(p.corner 1);\draw[blue] (p.corner 3)--(p.corner 8);
    \draw[blue] (p.corner 3)--(p.corner 6);\draw[blue] (p.corner 3)--(p.corner 5);

    \draw[blue] (p.corner 1)--(p.corner 8);
    \draw[blue] (p.corner 1)--(p.corner 6);\draw[blue] (p.corner 1)--(p.corner 5);

    \draw[blue] (p.corner 8)--(p.corner 6);
    \draw[blue] (p.corner 8)--(p.corner 4);\draw[blue] (p.corner 6)--(p.corner 4);
\end{tikzpicture}
\end{aligned}
\end{equation*}
    \caption{Kinematic shifts for two examples with $n=8$ ($4$ pions) and one with $n=9$ ($6$ pions).}
    \label{fig:mixed8pt2}
\end{figure}
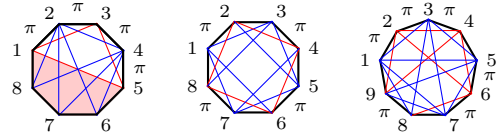

It is also straightforward to go beyond two pions. Let us first consider the special case with $\pi$'s all adjacent; for example, $[\pi_1\pi_2\pi_3\pi_4\Phi_5\Phi_6\Phi_7\Phi_8]$, the algorithm gives $-\delta_{1,3}=-\delta_{3,5}=-\delta_{1,5}=\delta_{2,4}=\delta_{2,i}=\delta_{4,i}$ for $i=6,7,8$ (see left of Fig.~\ref{fig:mixed8pt2}). In fact, the algorithm leads to a general formula for $2m$ adjacent $\pi$'s and $n-2m$ $\Phi$'s, $[\pi_1\cdots \pi_{2m}\Phi_{2m+1}\cdots\Phi_n]$. We can divide the $n$ vertices into three sets: odd, even for $\pi$'s and the remaining for $\Phi$'s, {\it i.e.} $o \in \{1, 3, \cdots, 2m{+}1\}$, $e\in \{2, 4, \cdots, 2m\}$, and $r\in \{2m{+}2, \cdots, n\}$, and the shifts are
\begin{equation}
\lim_{\delta\to \infty}\delta^{2m}  A_n^{\Phi^3}(-\delta_{o,o}=\delta_{e,e}=\delta_{e, r}=\delta)\,. 
\end{equation}
In particular, for odd-point amplitude with only 3 colored scalars, say $[\Phi_1\pi_2\pi_3\cdots \pi_{n-2}\Phi_{n-1}\Phi_n]$, the shifts take the familiar form: $-\delta_{e,e}=\delta_{o,o}=\delta$. Of course they are mixed amplitudes that appear in ``odd-odd" factorization near zero of pure NLSM amplitude~\cite{Zeros}. 

Let us now consider the $n=8$ ``alternating" case $[\pi_1 \Phi_2 \pi_3 \Phi_4 \pi_5 \Phi_6 \pi_7 \Phi_8]$, and the shifts are nicely given by the middle of Fig.~\ref{fig:mixed8pt2}. In fact, we can generalize this to the alternating case with $2m$ $\pi$'s and $2m$ $\Phi$'s, $[\pi_1\Phi_2\pi_3\Phi_4\cdots \pi_{4m-1} \Phi_{4m}]$. The shifts are
\begin{equation}
\begin{aligned}
&\delta_{o,o^{\prime}}=-\delta_{e,e^{\prime}}=\delta_{e,o}=\delta
\end{aligned}
\end{equation}
for odd (even) vertices, $o, o'$ ($e, e'$) where we only shift forbidden chords $(i j)$. It is easy to see that for $i$ odd, we must have $\lceil\frac{\abs{i-j}}{2}\rceil$ equal to some odd (positive) integer, and same for $\lfloor\frac{\abs{i-j}}{2}\rfloor$  for $i$ even. Only with these conditions the left/right lower amplitude has an odd number of $\pi$'s, and we need to shift such forbidden $X_{i,j}$ poles. 

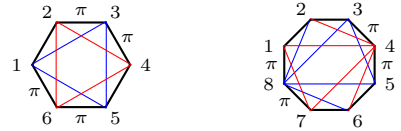
\begin{figure}
\centering
\begin{tikzpicture}[scale=1.1]
\node[regular polygon,minimum size = 1.3 cm,regular polygon sides=6,draw,thick] (p) at (0,0) {};
\draw[thick] (p.corner 1) node[above right=-2pt]{\scriptsize $3$};
\draw[thick] (p.corner 2) node[above left=-2pt]{\scriptsize $2$};
\draw[thick] (p.corner 3) node[left=0pt]{\scriptsize $1$};
\draw[thick] (p.corner 4) node[below left=-2pt]{\scriptsize $6$};
\draw[thick] (p.corner 5) node[below right=-2pt]{\scriptsize $5$};
\draw[thick] (p.corner 6) node[right=0pt]{\scriptsize $4$};
\node at (30:0.65) {\scriptsize $\pi$};
\node at (90:0.65) {\scriptsize $\pi$};
\node at (-90:0.65) {\scriptsize $\pi$};
\node at (-150:0.65) {\scriptsize $\pi$};

\draw[blue] (p.corner 1)--(p.corner 3);
\draw[blue] (p.corner 3)--(p.corner 5);
\draw[blue] (p.corner 1)--(p.corner 5);

\draw[red] (p.corner 2)--(p.corner 4);
\draw[red] (p.corner 6)--(p.corner 4);
\draw[red] (p.corner 2)--(p.corner 6);

\node[regular polygon,minimum size = 1.3 cm,regular polygon sides=8,draw,thick] (q) at (3,0) {};
\draw[thick] (q.corner 1) node[above right=-2pt]{\scriptsize $3$};
\draw[thick] (q.corner 2) node[above left=-2pt]{\scriptsize $2$};
\draw[thick] (q.corner 3) node[left=0pt]{\scriptsize $1$};
\draw[thick] (q.corner 4) node[left=0pt]{\scriptsize $8$};
\draw[thick] (q.corner 5) node[below left=-2pt]{\scriptsize $7$};
\draw[thick] (q.corner 6) node[below right=-2pt]{\scriptsize $6$};
\draw[thick] (q.corner 7) node[right=0pt]{\scriptsize $5$};
\draw[thick] (q.corner 8) node[right=0pt]{\scriptsize $4$};
\node at ($(q)+(0:0.7)$) {\scriptsize $\pi$};
\node at ($(q)+(40:0.7)$) {\scriptsize $\pi$};
\node at ($(q)+(180:0.7)$) {\scriptsize $\pi$};
\node at ($(q)+(220:0.7)$) {\scriptsize $\pi$};

\draw[blue] (q.corner 1)--(q.corner 4);\draw[blue] (q.corner 1)--(q.corner 7);
\draw[blue] (q.corner 7)--(q.corner 4);\draw[blue] (q.corner 4)--(q.corner 2);
\draw[blue] (q.corner 4)--(q.corner 6);
\draw[red] (q.corner 3)--(q.corner 5); \draw[red] (q.corner 3)--(q.corner 8);
\draw[red] (q.corner 5)--(q.corner 8); \draw[red] (q.corner 8)--(q.corner 2);
\draw[red] (q.corner 8)--(q.corner 6);

\end{tikzpicture}
    \caption{Blow up: from $[\Phi_1 \pi_2 \pi_3 \Phi_4 \pi_5 \pi_6]$ to $[(\Phi_1 \Phi_2) \pi_3 \pi_4 (\Phi_5 \Phi_6) \pi_7 \pi_8]$. }
    \label{fig:blowup}
\end{figure}

We note that this algorithm reproduces the cases $(i)$, where the shifts preserve skinny rectangle zeros for $[\Phi \cdots \Phi) (\pi \cdots \pi) (\Phi \cdots \Phi) (\pi \cdots \pi)]$. An example is shown in Fig.~\ref{fig:blowup}.

We finally give an example of an $n=9$ amplitude $[\Phi_1 \pi_2 \pi_3 \Phi_4 \pi_5 \pi_6 \Phi_7 \pi_8 \pi_9]$ with $6$ pions; the shifts are given in the right of Fig.~\ref{fig:mixed8pt2}.

\section{Loop-level mixed amplitudes}
\label{app:Loop}

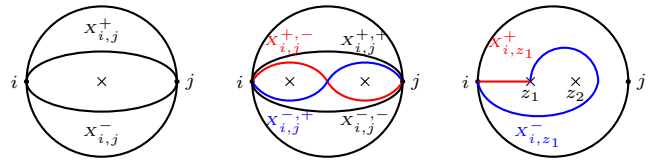
\begin{figure}\begin{tikzpicture}\draw[thick] (0,0) circle (1);
\draw (180:1) node[left=0pt]{\scriptsize $i$};
\draw (0:1) node[right=0pt]{\scriptsize $j$};
\draw (0,0.4) node[above=0pt]{\tiny $X_{i,j}^+$};
\draw (0,-0.4) node[below=0pt]{\tiny $X_{i,j}^-$};
\draw[thick,fill=black] (180:1) circle ( 0.02);
\draw[thick,fill=black] (0:1) circle ( 0.02);
\node[cross] (x1) at (0,0) {}; 	
\draw[thick] (0,0) ellipse (1 and 0.4 );

\draw[thick] (3,0) circle (1);
\draw (2,0) node[left=0pt]{\scriptsize $i$};
\draw (4,0) node[right=0pt]{\scriptsize $j$};
\draw[thick,fill=black] (2,0) circle ( 0.02);
\draw[thick,fill=black] (4,0) circle ( 0.02);
\node[cross] (y1) at (2.5,0) {}; 	
\node[cross] (y2) at (3.5,0) {}; 
\draw[thick,red] (2,0) to [bend left=60] (3,0);
\draw[thick,red] (3,0) to [bend right=60] (4,0);
\draw[thick,blue] (2,0) to [bend right=60] (3,0);
\draw[thick,blue] (3,0) to [bend left=60] (4,0);
\draw[thick] (3,0) ellipse (1 and 0.4 );
\draw (2.5,0.25) node[above=0pt]{\tiny \color{red}  $X_{i,j}^{+,-}$};
\draw (3.5,0.25) node[above=0pt]{\tiny  $X_{i,j}^{+,+}$};
\draw (2.5,-0.25) node[below=0pt]{\tiny \color{blue}  $X_{i,j}^{-,+}$};
\draw (3.5,-0.25) node[below=0pt]{\tiny  $X_{i,j}^{-,-}$};

\draw[thick] (6,0) circle (1);
\draw (5,0) node[left=0pt]{\scriptsize $i$};
\draw (7,0) node[right=0pt]{\scriptsize $j$};
\draw[thick,fill=black] (5,0) circle ( 0.02);
\draw[thick,fill=black] (7,0) circle ( 0.02);
\node[cross] (z1) at (5.7,0) {}; 	
\node[cross] (z2) at (6.3,0) {};
\draw[thick,red] (5,0)--(5.7,0);
\draw[thick,blue] (5,0) to [bend right=80] (6.6,0) arc (0:180:0.45);
\draw (5.45,0.2) node[above=0pt]{\tiny \color{red} $X_{i,z_1}^+$};
\draw (5.8,-0.4) node[below=0pt]{\tiny \color{blue} $X_{i,z_1}^-$};
\draw (z1) node[below=0pt]{\scriptsize $z_1$};
\draw (z2) node[below=0pt]{\scriptsize $z_2$};
\end{tikzpicture}
\caption{Loop variables for homologically inequivalent curves, which are identified for physical integrands.}
\label{fig:Loopvariables}
\end{figure}

\begin{figure}
\begin{tikzpicture}
    \node[regular polygon,minimum size = 1.5cm,regular polygon       sides=3,draw,thick] (p) at (0,0) {};
    \draw[thick] (p.corner 1) node[above=0pt]{\scriptsize $2$};
    \draw[thick] (p.corner 2) node[below=0pt]{\scriptsize $1$};
    \draw[thick] (p.corner 3) node[below=0pt]{\scriptsize $3$};
    \draw (0,0) node[cross] {};
    \draw[thick,red] (p.corner 1)--(0,0);
    \draw[thick,blue] (p.corner 2) .. controls (0,0.35) .. (p.corner 3);
    \draw[thick,red] (p.corner 1) .. controls (-45:0.35) .. (p.corner 2);
    \draw[thick,red] (p.corner 1) .. controls (-135:0.35) .. (p.corner 3);
    \draw [cyan,red,thick] plot [smooth, tension=1] coordinates { (p.corner 1) (-45:0.3) (-135:0.3) (p.corner 1)};
     \node at (270:0.6) {\scriptsize $\Phi$};
    \node at (150:0.6) {\scriptsize $\pi$};
    \node at (30:0.6) {\scriptsize $\pi$};
\end{tikzpicture}
    \caption{Kinematic shifts for $3$-point mixed amplitudes with $1,2$ to be pions.
    }
    \label{fig:mixeddeform}
\end{figure}
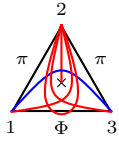

Let us present some preliminary results for loop-level mixed amplitudes. An important subtlety that we do not need for the pure NLSM case, is to carefully distinguish loop variables associated with homologically inequivalent curves (see Fig.~\ref{fig:Loopvariables}), {\it e.g.} at one-loop, $X_{i,j}^{\pm}$ are associated with curves going around the puncture in two different ways. For $A_{n,L}(\pi_1 \pi_2 \Phi_3 \cdots \Phi_n)$, we claim that the shifts are
\begin{equation}
X_{1,3}^{+,\dots,+}\to X_{1,3}^{+,\dots,+}-\delta, \quad X_{2,j}\to X_{2,j}+\delta,
\label{eq:shiftLoop}
\end{equation}
for $j=1,2,\ldots,n, z_{1},\ldots,z_{L}$. Here $X_{1,3}^{+,\dots,+}$ denotes the chord that bounds the $\Phi$ region while containing all the punctures. Since this factors on an amplitude with only a single $\Phi$, just as at tree-level, this chord must be shifted, and all $X_{2,j}$ variables are shifted oppositely. 

Applying the shifts \eqref{eq:shiftLoop} to the case where we have two $\pi$'s and a single $\Phi$ (see Fig.~\ref{fig:mixeddeform}) we get 
\begin{equation}
\begin{split}
    A_{3,1}&(\pi_1\pi_2\Phi_3)=\frac{1}{X_{1,z}}+\frac{1}{X_{3,z}}-\frac{X_{1,2}+X_{1,3}}{X_{1,1} X_{1,z}} \\
    &-\frac{X_{1,3}+X_{2,z}}{X_{1,z} X_{3,z}} -\frac{X_{1,3}+X_{2,3}}{X_{3,3} X_{3,z}}.
\end{split}
\end{equation}
In analogy to the pure pion cases, one can prove that mixed amplitudes obtained in this way have all the correct factorizations and single cuts.

More general configurations involve blowing up the $\pi$ regions and introducing new $\Phi$ regions. In both cases, we expect that an appropriate generalization of the tree-level rule should give us correct shifts, if they exist. It is plausible that in addition to shifting, we might also need to consider an average of shifted objects, for instance when multiple $\Phi$ regions coexist. 

\bibliographystyle{apsrev4-1.bst}
\bibliography{Refs.bib}

\end{document}